\documentclass[reprint, amsmath, amssymb, floatfix, aps]{revtex4-2}
\usepackage[utf8]{inputenc}
\usepackage{dcolumn}
\usepackage{graphicx}
\usepackage{xcolor}
\usepackage{float}
\usepackage{amsthm}
\begin{document}
\title{Particle image velocimetry analysis with simultaneous uncertainty quantification using Bayesian neural networks}
\author{Mia C Morrell}
\address{Los Alamos National Laboratory, XCP-8, Los Alamos, NM 87545, USA}
\author{Kyle Hickmann}
\address{Los Alamos National Laboratory, XCP-8, Los Alamos, NM 87545, USA}
\author{Brandon M. Wilson}
\address{Los Alamos National Laboratory, XCP-8, Los Alamos, NM 87545, USA}
\date{\today}
\begin{abstract}
    Particle image velocimetry (PIV) is an effective tool in experimental fluid mechanics to extract flow fields from images. Recently, convolutional neural networks (CNNs) have been used to perform PIV analysis with accuracy on par with classical methods. Here we extend the use of CNNs to analyze PIV data while providing simultaneous uncertainty quantification on the inferred flow field. The method we apply in this paper is a Bayesian convolutional neural network (BCNN) which learns distributions of the CNN weights through variational Bayes. We compare the performance of three different BCNN models. The first network estimates flow velocity from image interrogation regions only. Our second model learns to infer velocity from both the image interrogation regions and interrogation region cross-correlation maps. Finally, our best performing network derives velocities from interrogation region cross-correlation maps only. We find that BCNNs using interrogation region cross-correlation maps as inputs perform better than those using interrogation windows only as inputs and discuss reasons why this may be the case. Additionally, we test the best performing BCNN on a full test image pair, showing that $100\%$ of true particle displacements can be captured within its $95\%$ confidence interval. Finally, we show that BCNNs can be generalized to be used with multi-pass PIV algorithms with a moderate loss in accuracy, which may be overcome by future work on finetuning and training schemes. To our knowledge, this is the first effort to use Bayesian neural networks to perform particle image velocimetry.
\end{abstract}
\maketitle

\section{Introduction}

\emph{In this work, we present an alternate particle image velocimetry (PIV) uncertainty quantification method using Bayesian convolutional neural networks (BCNNs).} A data-driven uncertainty method, such as that outlined in this paper offers low-overhead, computationally inexpensive, algorithm independent uncertainty estimation.

\begin{figure*}
    \includegraphics{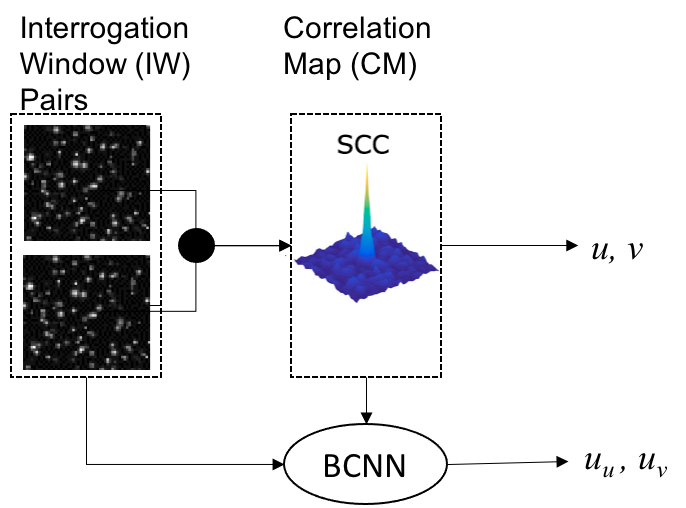}
    \caption{Sketch of data flow for the neural networks in this work (BCNNs and CNNs). Interrogation windows are specific regions of an image pair, and are used as inputs to some of our neural networks. The other possible input is the cross-correlation map of each interrogation window pair. In this work, we investigate using correlation maps only as inputs (CM), correlation maps and interrogation windows as inputs (CM+IW), and interrogation windows only as inputs (IW) to our neural networks. These neural networks output a prediction of particle displacement between members of a given image pair, indicated as $u_u, v_v$ in this sketch. We compare the results of each of these neural networks to that of OpenPIV, which uses correlation maps of interrogation window pairs to estimate the particle displacement $u, v$ between image frames. }
    \label{fig:PIV}
\end{figure*}

PIV is an imaging diagnostic used to determine fluid motion and has been applied to a variety of fluid dynamics applications (\emph{e.g.} turbulence \cite{Violato2011, Violato2013, Kahler2006, Lindken2002}, medical devices \cite{Hariharan2018, Kheradvar2012, Raghav2018}, atmospheric flow \cite{vanHout2007, Zhu2007, Kellnerova2012, Hu2011}, and nuclear reactors \cite{Yassin2008, Dominguez2009, Nguyen2018}). PIV is performed by acquiring successive images of small tracer particles seeded in the working fluid and illuminated by a thin laser sheet of the fluid (see Figure ~\ref{fig:PIV}). 

Classically, PIV algorithms infer the local fluid velocity from the correlation of consecutive image subregions called interrogation windows (IW). Full flow fields can then be determined for the entire image domain. Methods based on utilizing cross-correlation of consecutive images to estimate flow fields have been extremely successful in performing PIV analysis \cite{Adrian2005}. During the past 20 years, many algorithmic changes have improved the accuracy of calculated PIV velocities, including sub-pixel estimation, window shifting \cite{Wereley2002}, various correlation methods \cite{Meinhart2020, Westerweel1999, Sciacchitano2012, Eckstein2009}, and window deformation \cite{Huang1993,Scarano2001}.  

More recently, machine learning (ML) methods have been designed to replace the classical cross-correlation algorithm. Instead, convolutional neural networks (CNNs) applied to the image pair have been demonstrated to calculate the full flow field \cite{Cai2019_1, Cai2019_2, Rabault2017, Lee2017}. These CNN methods rival classical PIV methods in accuracy.

Whether using classical cross-correlation or advanced CNN methods, imaging conditions and algorithmic choices contribute to complex errors and uncertainties in PIV results. The prevalent use of PIV in research and industry has recently encouraged the development of many new methods of characterizing PIV errors and uncertainties. However, neither classical PIV methods nor CNN-based PIV methods have a generally accepted framework for uncertainty quantification (UQ). 

Multiple methods are found in the literature for quantifying PIV uncertainties \cite{Sciacchitano2013, Wieneke2015, Bhattacharya2018, Timmins2012, Smith2014, xue2015}. Strengths, limitations, and algorithms for these methods can be found in \cite{Sciacchitano2019}. In summary, the existing methods map uncertainty metrics from information contained in the image pairs (\emph{e.g.} particle image information, noise, etc.) or correlation map (\emph{e.g.} correlation distribution, intensity, noise, etc.) to a measure of confidence in the particle velocity prediction. The input-uncertainty mapping is informed empirically from synthetic data or analytically based on a limited number of predetermined features. 

In this work, we introduce a method of simultaneous PIV analysis and corresponding uncertainty quantification via deep neural network methods. Deep machine learning (DML) offers the possibility of learning complex, non-linear relationships between input and inferred velocity without the reliance on features designed through expert opinion. This flexibility of DML allows inference from a variety of features including predetermined features commonly used in classical PIV UQ and much more subtle features in the data. Our specific approach is to train Bayesian convolutional neural networks (BCNNs) to map correlation maps and image pairs to particle displacement.  The uncertainty prediction is then an inherent feature of the structure of the BCNN inference. To our knowledge, this work is the first instance of PIV analysis with automated UQ and the first application of Bayesian neural networks to the PIV task. BCNNs have been used successfully to identify complex mappings for other applications, including segmenting computed tomography (CT) scans, analyzing magnetic resonance imaging (MRI) data, detecting the location of light sources, and facial recognition \cite{Zafar2019, Zhao2018, Peretroukhin2017, LaBonte2020}. 

In the following, we describe the methodology of applying BCNNs to PIV uncertainty, including outlining the training data (Section~\ref{sec:training_data}) and providing an overview of convolutional and Bayesian neural networks applied to PIV (Section~\ref{sec:CNN}-\ref{sec:BCNN}). We then present the CNN and BCNN results for three inputs (Section~\ref{sec:results}): image pairs (IW), correlation map and image pairs (CM+IW), and correlation map (CM). Next, we present results applying a BCNN to a sequential image pair to show how BCNNs may be used in practice. Finally, we present results applying a BCNN to a multi-pass PIV algorithm in order to show how BCNNs can be used in conjunction with existing PIV algorithms (Section~\ref{sec:full}-\ref{sec:2pass}).

\section{Methodology}

\subsection{Synthetic PIV Image Training Data}
\label{sec:training_data}
In order to evaluate the performance of our BCNNs, we train and test BCNNs of identical architecture but different data inputs on the same training and testing data sets. Three network types are trained with the input type changing between each network. The different types of input studied are, 1) interrogation windows only (IW), 2) correlation maps and interrogation windows (CM+IW), and 3) correlation maps only (CM). The training data are $128 \times 128$ pixel artificial images which vary error-contributing parameters (particle displacement, particle diameter, particle density, shear, out-of-plane motion, and background noise). The parameters used to generate the training data are tabulated in Table~\ref{tbl:table_1}. Throughout this work, background noise level is represented by Gaussian white noise added to the synthetic image with variance which is a given fraction of the standard deviation of the original image. See Appendix~\ref{app:imagegen} for more information concerning synthetic image generation. 

We generate two test sets. The first test set (Test Set I) resembles the training data and is generated using the parameters tabulated in Table~\ref{tbl:table_1}. The second test set (Test Set II) varies each error-contributing parameter, one at a time, while holding all other error-contributing parameters at a constant value. The parameters used to generate Test Set II are shown in Table~\ref{tbl:table_2}. The purpose of Test Set I is to determine how well the BCNNs perform on data similar to their training data, while the purpose of Test Set II is to evaluate how the BCNNs respond to uncertainty introduced by variation in single error-contributing parameters.

\begin{table}
\centering
\caption{Parameters for training data and Test Set I}
\label{tbl:table_1}
  \begin{tabular}{ | p{5.5cm} | c | }
    \hline
    Parameter & Sample range \\ \hline
    \hline
    Diameter of particles (pixels) & $[1.0,5.0]$ \\ \hline
    Particle density (particles/square pixel) & $[0.012,0.117]$ \\ \hline
    Strength of shear (multiplier) & $[0.,0.02]$ \\ \hline
    Angle of shear (degrees) & $[0.,360.]$ \\ \hline
    x-displacement (pixels) & $[-4.,4.]$ \\ \hline
    y-displacement (pixels) & $[-4.,4.]$ \\ \hline
    Out-of-plane displacement (pixels) & $[0.,0.3]$ \\ \hline
    Noise level (fraction of standard deviation of image)& $0.0$ \\ \hline
  \end{tabular}
\end{table}

\begin{table}
\centering
\caption{Parameters used to generate Test Set II}
\label{tbl:table_2}
  \begin{tabular}{ | p{5.5cm} | p{1.cm} | p{2cm} |}
    \hline
    Parameter & Default value & sweep range/value\\ \hline
    \hline
    Diameter of particles (pixels) & $3.0$ &  $[0.5, 5.]$ \\ \hline
    Particle density (particles/square pixel) & $0.098$ & $[0.0098, 0.146]$ \\ \hline
    Strength of shear  (multiplier) & $0.0$ &  $[0.0,0.1]$ \\ \hline
    Angle of shear (degrees) & $0.0$ & $0.0$ \\ \hline
    x-displacement (pixels) & $-1.0$ & $[-5.20, 5.20]$ \\ \hline
    y-displacement (pixels) & $1.0$ & $[-5.20, 5.20]$ \\ \hline
    Out-of-plane displacement (pixels) & $0.0$ & $[0.0, 0.4]$ \\ \hline
    Noise level (fraction of standard deviation of image) & $0.0$ & $[0.0, 0.2]$ \\ \hline
  \end{tabular}
\end{table}

\subsection{Convolutional Neural Network}
\label{sec:CNN}
\begin{figure*}
    \includegraphics{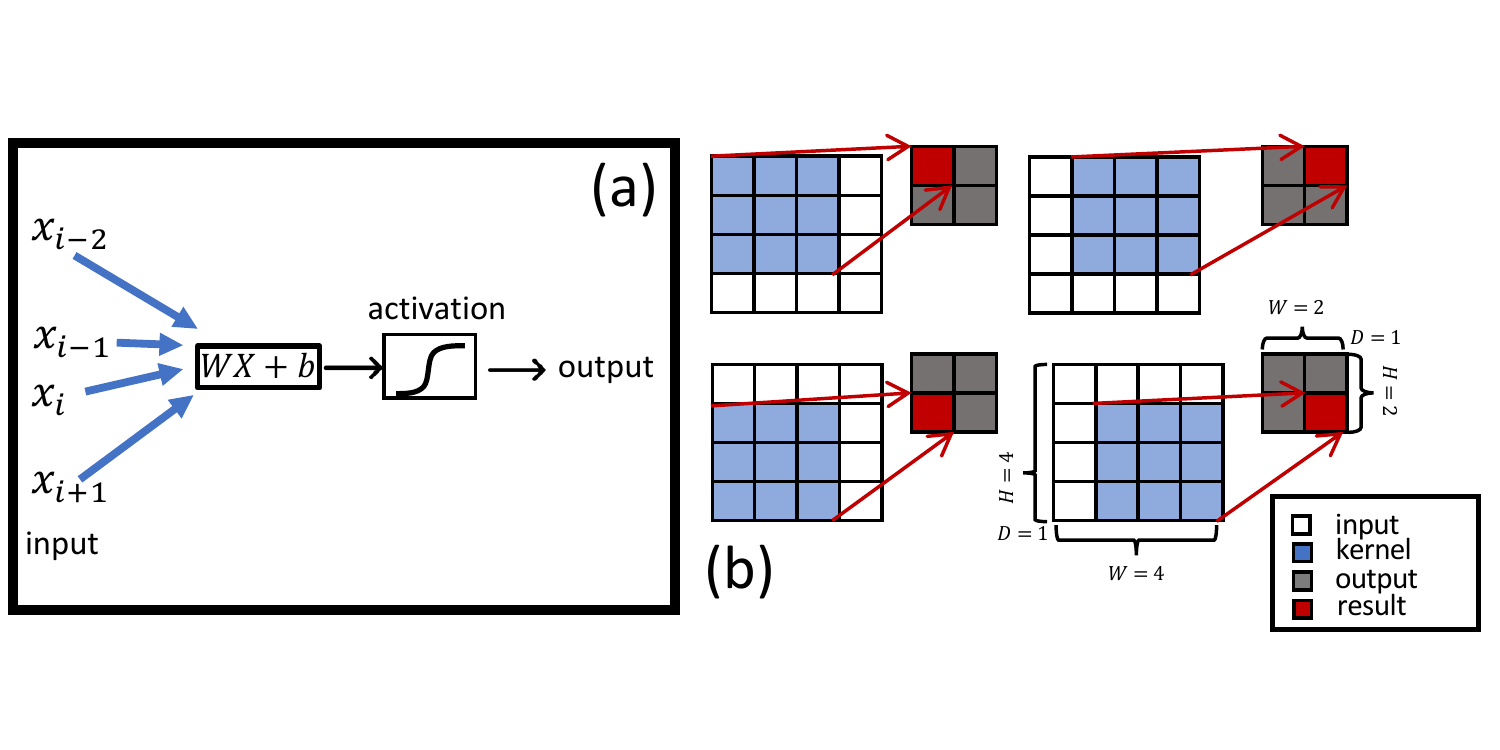}
    \caption{\textbf{(A)} Basic operation of a single neural network layer. Inputs $X$ to the neural network undergo a linear transformation $WX+b$ (where $W$ and $b$ are scalar matrices referred to as \emph{weights} and \emph{biases}, respectively) and are passed through a nonlinear function, or \emph{activation} to produce an output. \textbf{(B)} A sketch of the operation of a convolutional layer with a single filter of kernel size $3 \times 3$ with stride of 1. \emph{Stride} indicates how far the kernel slides between each step of the convolution. The kernel applies a transformation to the shaded blue area, outputting a convolutional result, denoted in red for each step of the convolution. }
    \label{fig:cnn}
\end{figure*}

\begin{figure*}
    \includegraphics{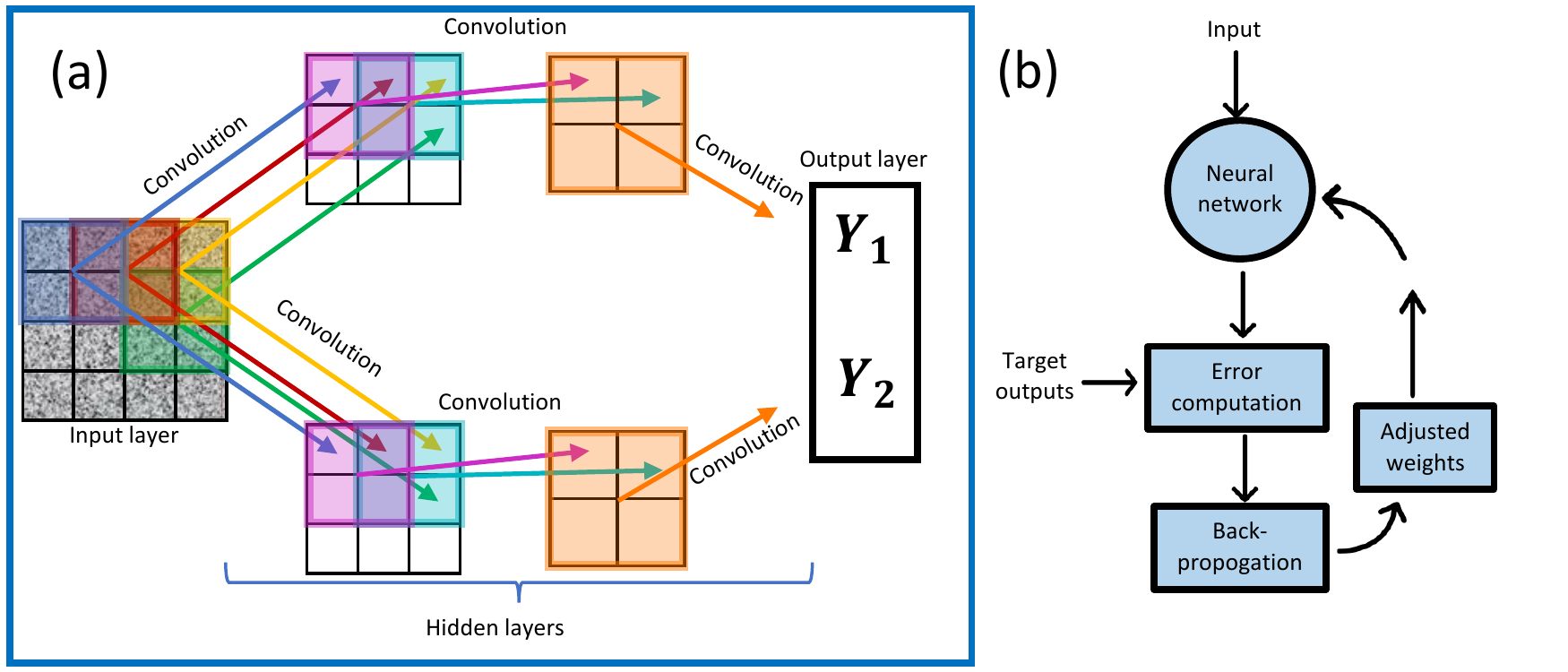}
    \caption{\textbf{(A)} Sketch of a convolutional neural network (CNN). At the input layer, image data is fed into the CNN. Within the hidden layers, the image passes through convolutional layers, the operation of which is described in Figure~\ref{fig:cnn}. In this sketch, the image passes through 3 convolutional layers. The first, second, and third convolutional layers all have 2 filters with kernel size of $2 \times 2$ and stride of 1. Transparent squares represent filter locations on each layer input. Each color square represents a different filter location. After the final convolutional layer, an output of 2 vectors $Y_1$ and $Y_2$ is obtained at the output layer. \textbf{(B)} Diagram showing training process for a neural network. Inputs are fed into the neural network, producing predictions. The error of these predictions is calculated. Via backpropogation, an algorithm which computes the neural network's gradient in weight space with respect to a loss function, the weights of the neural network (illustrated in Figure~\ref{fig:cnn}A) are updated. This process continues, minimizing the error between the predicted and target outputs via the loss function.}
    \label{fig:cnn_layers}
\end{figure*}

Convolutional neural networks (CNNs) have been widely used in computer vision tasks, including that of PIV \cite{Cai2019_1, Cai2019_2, Rabault2017, Lee2017}. CNNs are a type of deep neural network designed to exploit structures such as locality and translational invariance found in images \cite{Lecun1995}. The basic building block of deep neural networks (including CNNs) is a \emph{neuron} $a_i$ mapping $\mathbb{R}^n$ to $\mathbb{R}^m$ which takes in an input vector $\mathbf{X} \in \mathbb{R}^n$ and outputs a non-linearly transformed vector, $\mathbf{Y} = a_i(\mathbf{X}) = \sigma_i(\mathbf{W}\mathbf{X}+\mathbf{b})$ in $\mathbb{R}^m$ (see Figure~\ref{fig:cnn}). Here $\sigma_i$ is a nonlinear function applied elementwise, $\mathbf{b}$ is a vector of biases, and $\mathbf{W}$ is a weight matrix. Neurons of the network are arranged into layers in which the output of one layer is the input of the following layer. The first layer of neurons processes the input directly, the middle layers are the hidden layers, and the last layer is the output layer which returns the prediction. A \emph{convolutional} neural network replaces the weight matrix $\mathbf{W}$ with a convolutional filter, say $\mathbf{w}$, so that the neurons can be expressed as $\mathbf{Y} = \sigma_i(\mathbf{w} \ast \mathbf{X}+\mathbf{b})$. A sketch of input, hidden, and output layers in a CNN is shown in Figure~\ref{fig:cnn_layers}A.

The use of convolutional filters in CNNs has two main advantages. First, it drastically reduces the number of trainable parameters in the network and second, it creates translation invariant nonlinear filters in the network allowing the network to learn feature structure independent of where a feature occurs in the image.

\begin{figure}
    \includegraphics{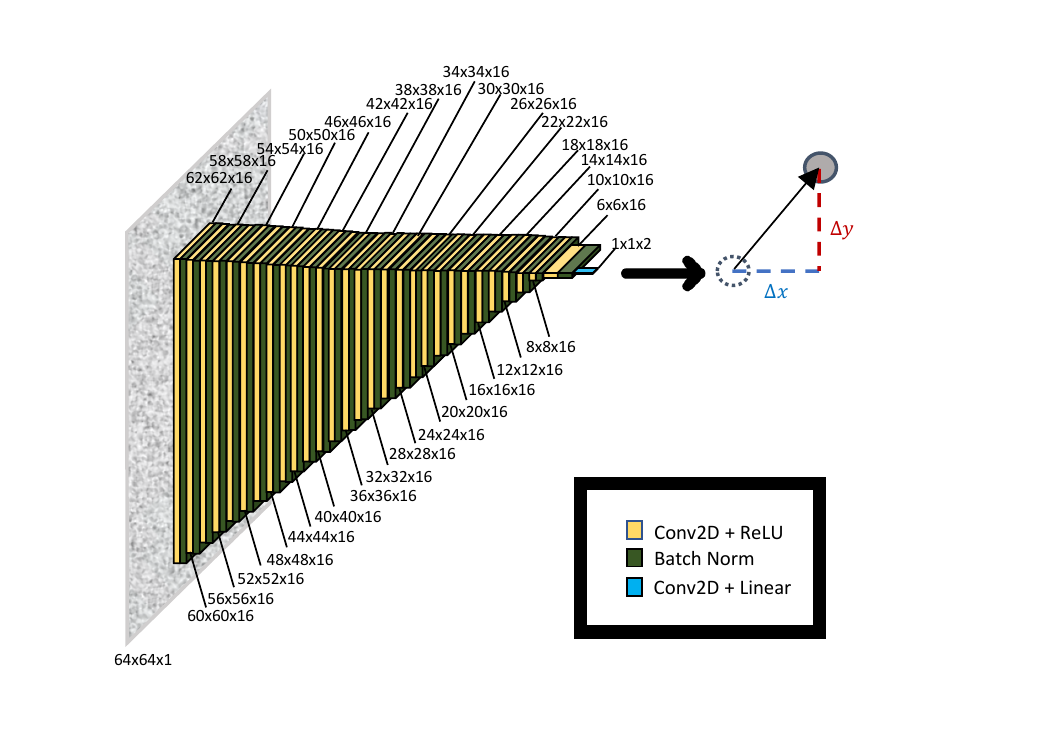}
    \caption{Output shape at each layer of the neural networks with correlation maps only used as inputs (CM). Similar diagrams for the neural networks using interrogation windows only as inputs (IW) and neural networks using both correlation maps and interrogation windows as inputs (CM+IW) can be found in the Appendix~\ref{app:appendix} (Figure~\ref{fig:iw-bcnn} and Figure~\ref{fig:cm+iw-bcnn}, respectively). All neural networks follow the same basic structure: repeated application of convolutional layers followed by batch normalization layers until the input is decreased in dimension to that of a 2D vector.}
    \label{fig:cm-bcnn}
\end{figure}

 Input to the CNN (IW, IW+CM, or CM in this paper) is fed into the input layer and passed through each layer of the CNN yielding a prediction of the particle displacement vectors between the interrogation window pairs at the output layer, as shown in Figure~\ref{fig:cm-bcnn}. In this work, we follow the approach of Reference~\cite{Springenberg2015} and include only convolutional and batch normalization layers in our CNNs. In the following, we provide a brief overview of each of these layers. See \cite{Mehta2019} for a more detailed introduction to CNNs and deep neural networks in general.

\subsubsection{Convolutional layers. ---} 
Convolutional layers apply a set of convolutional filters to the input in each layer. A 2D convolutional layer consists of neurons arranged in 3-dimensional blocks and is specified by its height, width, and depth. Each of these neurons computes an output value from its input in the previous layer. The height and width indicate the height and width of the layer in the spatial plane (in neurons), while the depth indicates the number of filters in the layer. Neurons which belong to a given filter share weights and biases. Filters use a small window of the previous layer as inputs, and the convolution is computed by running these filters over the entire spatial plane, as shown in Figure~\ref{fig:cnn}B. In this work, we use only linear, $\sigma_i(x) = x$, and \emph{rectified linear unit} (ReLU), $\sigma_i(x) = \max(0, x)$, activations.

\subsubsection{Batch normalization layers. ---} 
Batch normalization, a widely used regularization scheme for neural networks, scales and shifts the output of a neural network layer such that the result has approximately zero mean and unit variance \cite{Ioffe2015}. Batch normalization is used to increase the speed and stability of neural networks. Although batch normalization was originally recommended for use after convolutional layers and before activations, in this work we apply batch normalization layers \emph{after} activations. We use this order of convolutional layers, activations, and batch normalization since it produces more stable neural networks in the case of our chosen problem and network architecture. 

\subsubsection{Training. ---}
\label{sec:CNN:training}
During training, the CNN learns by iteratively optimizing its biases and weights in order to minimize a loss function that quantifies how well the network fits the training data. Optimizers used are usually based off of gradient descent algorithms with the key difference that each evaluation of the loss function is only performed on a small subset, known as a \emph{batch}, of the full training set. After each evaluation of the loss function and performance of the gradient descent step a new batch is chosen. This allows the network to learn from huge training sets but also implies that the loss function being reduced changes from step to step of the optimization. This type of optimization problem is referred to as a \emph{stochastic} optimization. 

Gradients of the loss are computed through a \emph{forward pass} and a \emph{backward pass} algorithm. During the forward pass, inputs are propogated through the CNN and outputs are returned. During the backward pass gradients are computed using the chain rule, the loss function is evaluated, and neuron weights are updated using gradient descent. \emph{Backpropogation}, an algorithm which computes the neural network's gradient in weight space with respect to a loss function, is a key tool used during the backward pass \cite{Rumelhart1986}. See Figure~\ref{fig:cnn_layers}B for an illustration of the training process.

In this work, we minimize the mean squared error loss function (MSE) using the Adam optimizer \cite{Kingma2014} while training our CNNs. This has become a standard training procedure for deep neural networks.

\subsubsection{CNN structure. ---} 
In this paper, for an input size of $32 \times 32$ pixels, we apply 13 convolutional layers with 16 filters each and a kernel size of $3 \times 3$ pixels followed by a convolutional layer with 2 filters and kernel size of $6 \times 6$. Between each convolutional layer we apply an activation, either ReLU or linear, and after each activation we apply a batch normalization layer (except for the last activation). See Figure~\ref{fig:cm-bcnn} for an example diagram of the CM CNN. Complete diagrams of the IW and CM+IW CNNs can be found in Figure~\ref{fig:iw-bcnn}, Figure~\ref{fig:cm+iw-bcnn} of the Appendix.

\subsection{Bayesian Neural Network}
\label{sec:BCNN}
Bayesian neural networks offer a solution to the deterministic neural network's lack of pointwise uncertainty quantification. In this work, we follow the approach of Reference~\cite{Blundell2015} to Bayesian neural networks. Bayesian neural networks account for the uncertainty of neural network weights by estimating the posterior prediction distribution 
\begin{equation}{\label{eq:posterior}}
    p(y|\textbf{x}, \mathcal{D}) = \int p(y| \textbf{x}, \textbf{w})p(\textbf{w}|\mathcal{D})d\textbf{w}
\end{equation}
where $\mathbf{x}$ is the input to the neural network with label $y$, $\mathcal{D} = \{(\mathbf{x}_i, y_i)\}$ is the training data, $\textbf{w}$ are the weights of the neural network, and $p(\mathcal{D}|\mathbf{w}) = \prod_i p(y_i|\mathbf{x}_i, \mathbf{w}_i)$ is the likelihood function, or the probability of predicting $\{y_i\}$ given inputs $\{\mathbf{x}_i\}$ and weights $\{\mathbf{w}_i\}$. Thus for a given input $\hat{\textbf{x}}$, a Bayesian neural network returns the probability of observing its corresponding label $\hat{y}$,
\begin{equation}
    p(\hat{y}|\hat{\textbf{x}}) = \mathbb{E}_{p(\mathbf{w}|\mathcal{D})}p(\hat{y}|\hat{\mathbf{x}}, \mathbf{w}).
\end{equation}

\subsubsection{Deriving a loss function}
In order to approximate the true posterior $p(\textbf{w}|\mathcal{D})$, we must apply variational inference by minimizing the Kullback-Leibler (KL) divergence between the true posterior $p(\textbf{w}|\mathcal{D})$ and a distribution of known form $q_{\boldsymbol{\theta}}(\textbf{w})$. We do this by minimizing the variational free energy as a function of $\boldsymbol(\theta)$, 
\begin{equation}{\label{eq:cost}}
    \mathcal{F}(\mathcal{D}, \boldsymbol{\theta})  = \text{KL}(q_{\boldsymbol{\theta}}(\mathbf{w})||\,p(\mathbf{w})) - \mathbb{E}_{q_{\boldsymbol{\theta}}(\mathbf{w})} \log p(\mathcal{D}|\mathbf{w})
\end{equation}
where the first term is the KL divergence between the variational posterior $q_{\boldsymbol{\theta}}(\mathbf{w})$ and the prior distribution on the weights, $p(\mathbf{w})$, and the second term is the expected value of the log-likelihood of data $\mathcal{D}$ given weights $\mathbf{w}$ distributed by the variational posterior. We can approximate Eq.~\ref{eq:cost} via sampling $\mathbf{w}^{(j)}$ from $q_{\boldsymbol{\theta}}(\mathbf{w})$, 
\begin{equation}{\label{eq:costapprox}}
\mathcal{F}(\mathcal{D}, \boldsymbol{\theta})  \approx \frac{1}{N}\sum_{j=1}^{N}[ \log q_{\boldsymbol{\theta}}(\mathbf{w}^{(j)})-\log p(\mathbf{w}^{(j)}) - \log p(\mathcal{D}|\mathbf{w}^{(j)})].
\end{equation}

Since we are performing regression, we can substitute the mean squared error function for the negative log likelihood $- \frac{1}{N}\sum_{j=1}^{N} \log p(\mathcal{D}|\mathbf{w}^{(j)})$, assuming Gaussian $ p(y_i|\mathbf{x}_i, \mathbf{w}_i)$ with fixed standard deviation (see Appendix~\ref{app:math}).

In this work, we use take $\boldsymbol{\theta} = (\boldsymbol{\mu}, \boldsymbol{\sigma})$ and approximate the true posterior $p(\mathbf{w} | \mathcal{D})$ with a Gaussian $q_{\boldsymbol{\theta}}(\mathbf{w})$ where $\boldsymbol{\mu}$ is the mean and $\boldsymbol{\sigma}$ is the standard deviation. Therefore, compared to a deterministic neural network of the same structure, our Bayesian neural network has twice the number of parameters.

\subsubsection{Training}
Training iterations consist of a forward and a backward pass, as in deterministic neural networks (described previously in \ref{sec:CNN:training}). The forward pass is performed by drawing a sample from $q_{\boldsymbol{\theta}}(\mathbf{w})$, which is then used to evaluate Eq.~\ref{eq:cost}. The backward pass updates values of weight distribution parameters $\boldsymbol{\mu}$ and $\boldsymbol{\sigma}$ via backpropogation, which is made possible by the Flipout Monte Carlo estimator \cite{Wen2018}. 

\subsubsection{Predicting particle displacement vectors}
A Bayesian neural network effectively constructs an ensemble of neural networks. By sampling this network ensemble, we obtain an ensemble of outputs for a single input from which statistics such as mean and standard deviation can be computed. To make predictions using the Bayesian neural network, we draw a sample from $q_{\boldsymbol{\theta}}(\mathbf{w})$ and use it to evaluate the output of the neural network. For any given input, we draw 2000 samples from $q_{\boldsymbol{\theta}}(\mathbf{w})$ and compute the mean and standard deviation of the resulting 2000 neural network outputs. We therefore can capture the center and spread of each prediction made by the Bayesian neural network.

\subsubsection{BCNN structure}
To perform PIV using a BCNN, we use the exact neural network structures and inputs used in our CNNs (shown in Figure~\ref{fig:iw-bcnn}, Figure~\ref{fig:cm+iw-bcnn}, and Figure~\ref{fig:cm-bcnn}). As described in the context of our CNNs, the inputs to the BCNN are either 1) interrogation window pairs (IW), 2) interrogation window pairs and their correlation maps (CM+IW), or 3) the cross-correlation of interrogation window pairs (CM). This input is fed into the BCNN, producing a prediction of the particle displacement vectors between the interrogation window pairs at the output layer. 

Our BCNNs differ from our CNNs since they use 2D convolutional variational layers instead of 2D convolutional layers. Convolutional variational layers differ from convolutional layers, as convolutional variational layers use distributions for weights as explained in the previous section. For all convolutional variational layers, the Flipout algorithm \cite{Wen2018} is used to estimate gradients for backpropagation and each layer is initialized with the standard normal prior.

\section{Results}

\subsection{BCNN and CNN performance}
\label{sec:results}

Here we demonstrate that the BCNN and CNN using the correlation maps only as inputs (CM-BCNN and CM-CNN) perform the best on both Test Set I and Test Set II out of the three BCNNs and three CNNs tested. We show that the CM-BCNN is capable of matching or exceeding a simple single pass PIV algorithm implemented using OpenPIV \cite{OpenPIVmatlab, OpenPIV} in accuracy during sweeps of many of the tested error-contributing parameters. Note that in this work, we use OpenPIV without any of its error-correction or window deformation capabilities for a fair comparison to and preliminary demonstration of BCNNs applied to PIV. For use with error-correction or window deformation, inputs generated from these advanced algorithms can be incorporated into the BCNN training following the methods outlined in this paper. These results are followed by a discussion of the relation between the tested BCNNs, their corresponding CNNs, and the OpenPIV algorithm. Next, we perform full PIV on an entire image using the best performing network, the CM-BCNN. Finally, we implement a ``double pass'' algorithm using the CM-BCNN to evaluate generalizability to correlation maps generated via slightly different algorithms.

\subsubsection{Test Set I}
\begin{figure*}
    \includegraphics{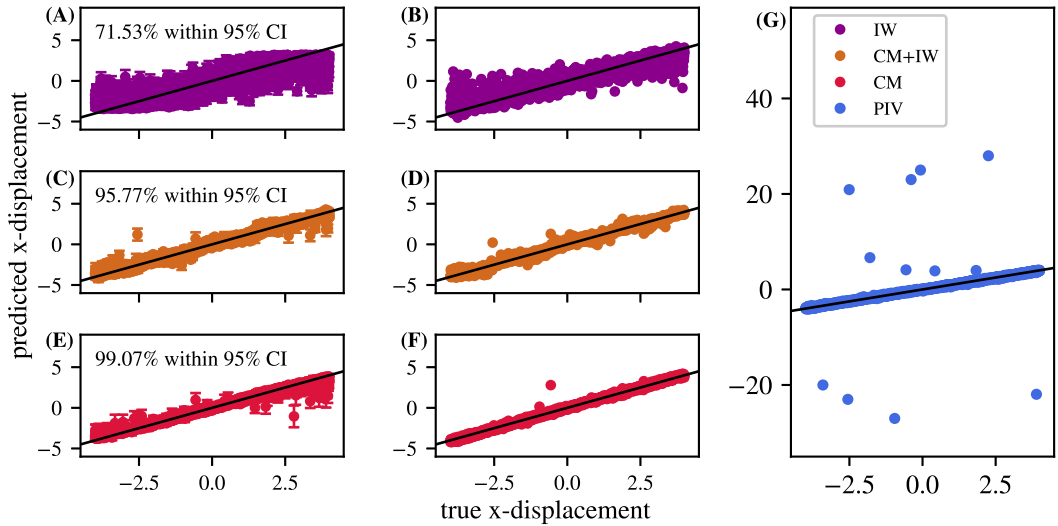}
    \caption{Predicted particle displacements versus true displacements in the $x$-direction. Predicted versus true particle displacement in the $y$-direction are similar and are shown in Appendix~\ref{app:appendix} (Figure~\ref{fig:testset1_y}). \textbf{(A, C, E)} show particle displacements predicted by Bayesian neural networks with interrogation windows only as inputs (IW-BCNN, purple), correlation maps and interrogation windows as inputs (CM+IW-BCNN, brown), and correlation maps only as inputs (CM-BCNN, red), respectively, versus true particle displacements. Error bars are standard deviations. \textbf{(B, D, F)} show particle displacements predicted by the deterministic versions of \textbf{(A, C, E)} (IW-CNN, CM+IW-BCNN, CM-BCNN), respectively versus true particle displacements. \textbf{(F)} shows particle displacements predicted by OpenPIV (PIV, blue) versus true particle displacements.  We observe that the CM-BCNN and the CM-CNN perform the best out of the BCNNs and CNNs tested, respectively. All of the tested BCNNs and CNNs decrease the number of large errors in displacement, or spurious displacements, compared to the OpenPIV algorithm \textbf{(F)}.}
    \label{fig:testset1_x}
\end{figure*}

First, we compare how well our BCNNs learn to extract displacement vectors from interrogation window and/or correlation map information by evaluating their performance on Test Set I, a data set similar to the training set. Figure~\ref{fig:testset1_x} shows the performance of the BCNN (left column), CNN (center column), and OpenPIV algorithm (right column) in predicting the $x$ displacement. Predictions of $y$ displacement are similar to those shown in Figure~\ref{fig:testset1_x} and are shown in Appendix~\ref{app:appendix} (Figure~\ref{fig:testset1_y}). The three network types, varying input classes, are also compared (IW (top), CM+IW (middle), and CM (bottom)).

The BCNN and CNN models trained on CM+IW (CM+IW-BCNN, CM+IW-CNN) and CM (CM-BCNN, CM-CNN) show similar performance in predicting displacements to OpenPIV. However, both BCNN and CNN methods predict fewer spurious vectors, or particle displacement predictions with large errors, than the OpenPIV algorithm. This suggests BCNN and CNN methods may reduce the need for complex spurious vector detection algorithms. Additionally, BCNN and CNN methods may provide a rigorous means to identify and replace spurious vectors.

An advantage of the BCNN method is the accompanying predicted uncertainties. Predicted uncertainties (95\% confidence intervals) capture $\approx72\%$, $\approx96\%$ and $\approx99\%$ of the true displacements for the IW, CM+IW and CM training sets, respectively. We see that the BCNN using the correlation maps only as inputs (CM-BCNN) performs the best out of the three BCNNs tested on Test Set I. This is most likely due to the additional feature extraction that cross-correlating the interrogation window pairs provides, and the omission of potentially uninformative information originating from the raw interrogation window pair.

Here we note some interesting phenomena that can be observed in Figure~\ref{fig:testset1_x}. First, we see that the BCNN and CNN predictions are inferior when trained on IW input alone. However, the BCNN predicted uncertainties (95\% confidence intervals) capture $\approx72\%$ of the true displacements, which is a surprisingly significant proportion given that no feature extraction (i.e. cross-correlation) has been applied. While, in this work, we have found poor performance using CNN architectures on IW input, it is always possible that variations in training hyperparameters or network architecture could allow for improved performance.

Additionally, upon inspection, both the CM+IW-CNN and CM+IW-BCNN networks may be impacted by pixel-locking phenomena. This is indicated by clustering of increased errors near integer true displacements. This may be due to certain patterns of operations learned by both the CM+IW-BCNN and CM+IW-CNN in the interrogation window features. Further investigation of the \emph{activations} of these networks, or the result of each network layer, may allow us to discern whether the CM+IW-BCNN and CM+IW-CNN are learning patterns which contribute to pixel-locking. This is outside the scope of the present report but will be explored in future work. 

Furthermore, we observe that the BCNNs seem to predict larger uncertainties and/or predict less accurate predictions at the boundaries of their training regime ($[-4, 4]$ pixels). This may be due to two causes. First, it may be more difficult for the BCNNs to predict larger displacements, a trend consistent with that of traditional PIV methods. Second, BCNNs may be more sensitive to the training data limits. If the latter were true, training the BCNN on a wider range of displacements would improve the accuracy away from the training data boundaries.

Finally, we note that the results displayed by each BCNN and its corresponding CNN (i.e. CM-BCNN and its deterministic version, CM-CNN) in Figure~\ref{fig:testset1_x} are quantitatively similar. This similarity emerges because BCNNs are effectively ensembles of CNNs with identical architectures. Thus, any appropriately trained CNN should return predictions contained within the prediction distribution of an appropriately trained BCNN of identical architecture. We observe this behavior in Figure~\ref{fig:testset1_x}, and additionally in Figure~\ref{fig:displacement}-\ref{fig:noise_level}, which are discussed in the next section. 

\subsubsection{Test Set II}
\label{sec:testsetii}
Next, we evaluate how each BCNN responds to changes in error contributing parameters by performing predictions on Test Set II, a data set which varies one error contributing parameter at a time while holding all other error contributing parameters constant. 

\begin{figure*}
    \includegraphics{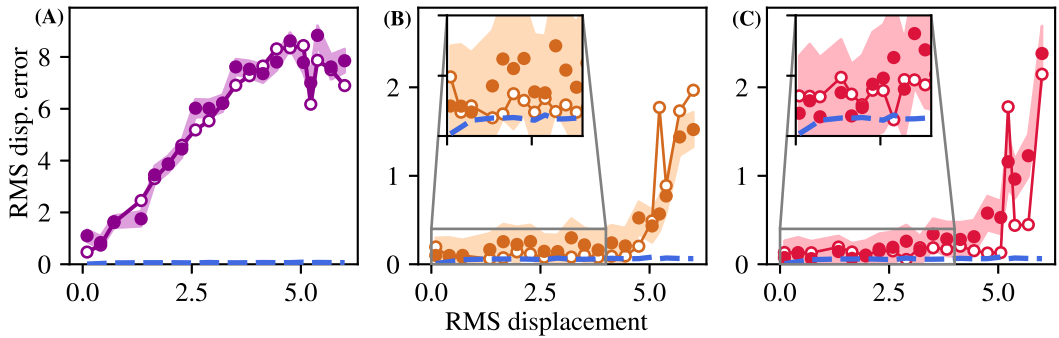}
    \caption{RMS error of particle displacements predicted by BCNNs and CNNs using interrogation windows only (IW, \textbf{A}), correlation maps and interrogation windows (CM+IW, \textbf{B}), and correlation maps only as input (CM, \textbf{C}) versus RMS particle displacement. Results from BCNNs are marked in solid circles with shaded regions representing standard deviations about the mean particle displacement prediction, while results from CNNs are marked in open circles. Each data point, solid or open, is an average of 100 samples. Shaded regions are standard deviations predicted by the BCNN, averaged over 100 samples. Blue dashed lines are the RMS error of predictions made by OpenPIV. We observe that as in Figure~\ref{fig:testset1_x}, the CM-BCNN and the CM-CNN (\textbf{C}) perform best out of the BCNNs and CNNs tested, respectively.  }
    \label{fig:displacement}
\end{figure*}

\begin{figure*}
    \includegraphics{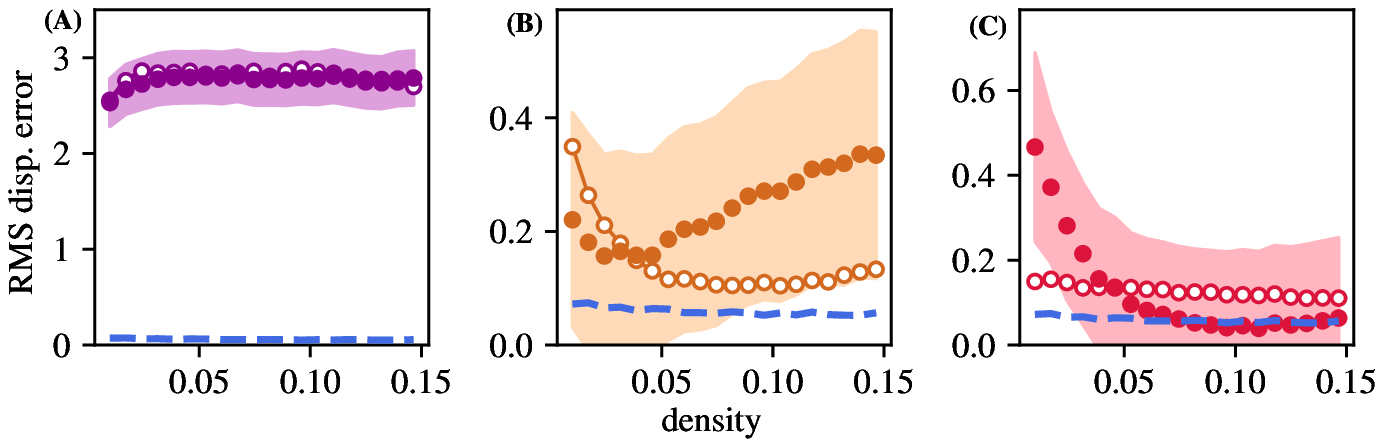}
    \caption{RMS error of particle displacements predicted by BCNNs and CNNs using interrogation windows only (IW, \textbf{A}), correlation maps and interrogation windows (CM+IW, \textbf{B}), and correlation maps only as input (CM, \textbf{C}) versus particle density. Results from BCNNs are marked in solid circles with shaded regions representing standard deviations about the mean displacement prediction, while results from CNNs are marked in open circles. Each data point, solid or open, is an average of 100 samples. Shaded regions are standard deviations predicted by the BCNN, averaged over 100 samples. Blue dashed lines are the RMS error of predictions made by OpenPIV. We observe that the CM-BCNN and the CM-CNN (\textbf{C}) perform best out of the BCNNs and CNNs tested, respectively.}
    \label{fig:density}
\end{figure*}

\begin{figure*}
    \includegraphics{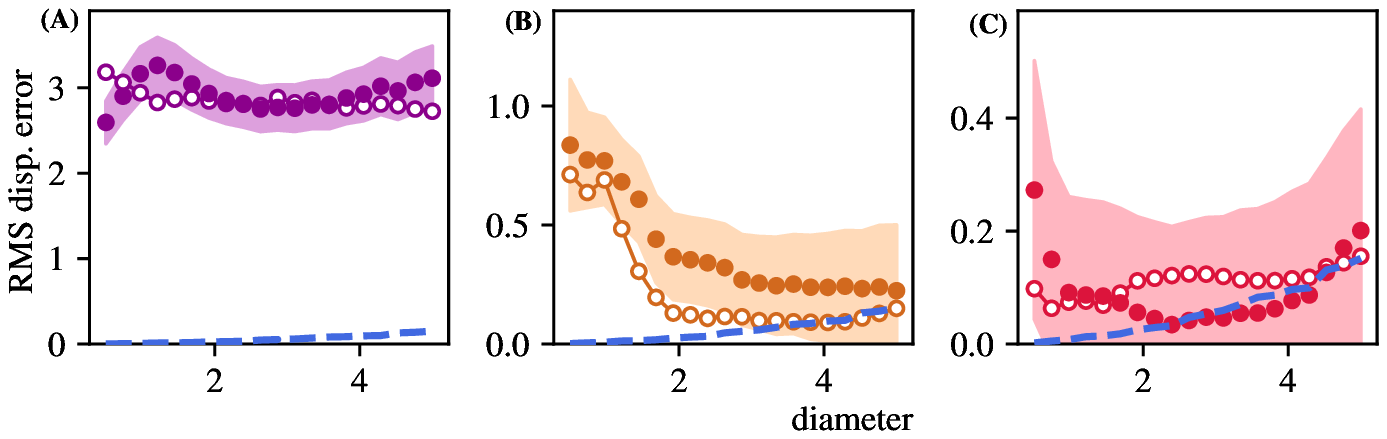}
    \caption{RMS error of particle displacements predicted by BCNNs and CNNs using interrogation windows only (IW, \textbf{A}), correlation maps and interrogation windows (CM+IW, \textbf{B}), and correlation maps only as input (CM, \textbf{C}) versus particle diameter. Results from BCNNs are marked in solid circles with shaded regions representing standard deviations about the mean displacement prediction, while results from CNNs are marked in open circles. Each data point, solid or open, is an average of 100 samples. Shaded regions are standard deviations predicted by the BCNN, averaged over 100 samples. Blue dashed lines are the RMS error of predictions made by OpenPIV. We observe that as in Figure~\ref{fig:testset1_x}, the CM-BCNN and the CM-CNN (\textbf{C}) perform best out of the BCNNs and CNNs tested, respectively.}
    \label{fig:diameter}
\end{figure*}

\begin{figure*}
    \includegraphics{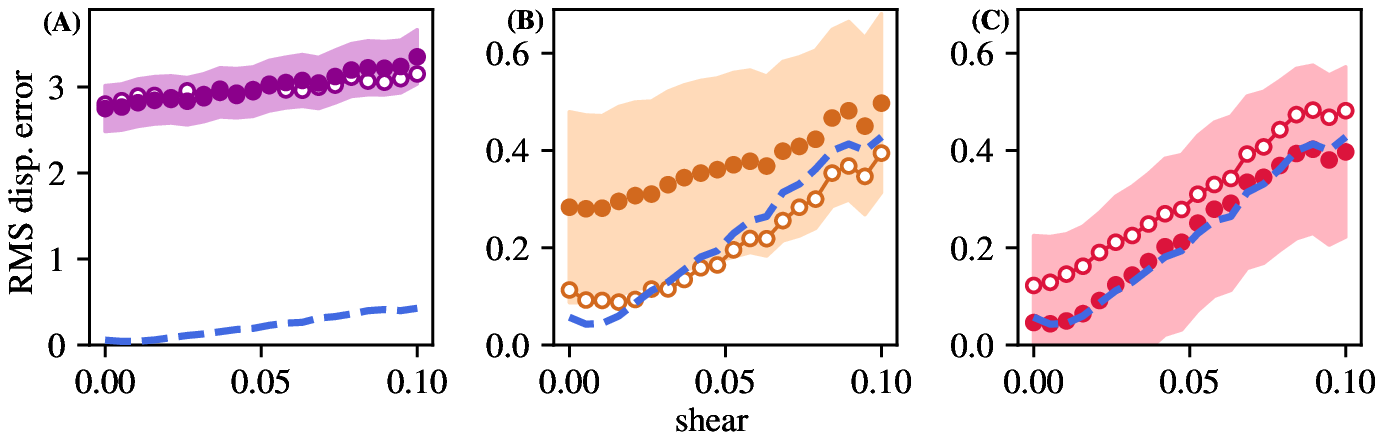}
    \caption{RMS error of particle displacements predicted by BCNNs and CNNs using interrogation windows only (IW, \textbf{A}), correlation maps and interrogation windows (CM+IW, \textbf{B}), and correlation maps only as input (CM, \textbf{C}) versus particle shear. Results from BCNNs are marked in solid circles with shaded regions representing standard deviations about the mean displacement prediction, while results from CNNs are marked in open circles. Each data point, solid or open, is an average of 100 samples. Shaded regions are standard deviations predicted by the BCNN, averaged over 100 samples. Blue dashed lines are the RMS error of predictions made by OpenPIV. We observe that as in Figure~\ref{fig:testset1_x}, the CM-BCNN and the CM-CNN (\textbf{C}) perform best out of the BCNNs and CNNs tested, respectively.}
    \label{fig:shear}
\end{figure*}

\begin{figure*}
    \includegraphics{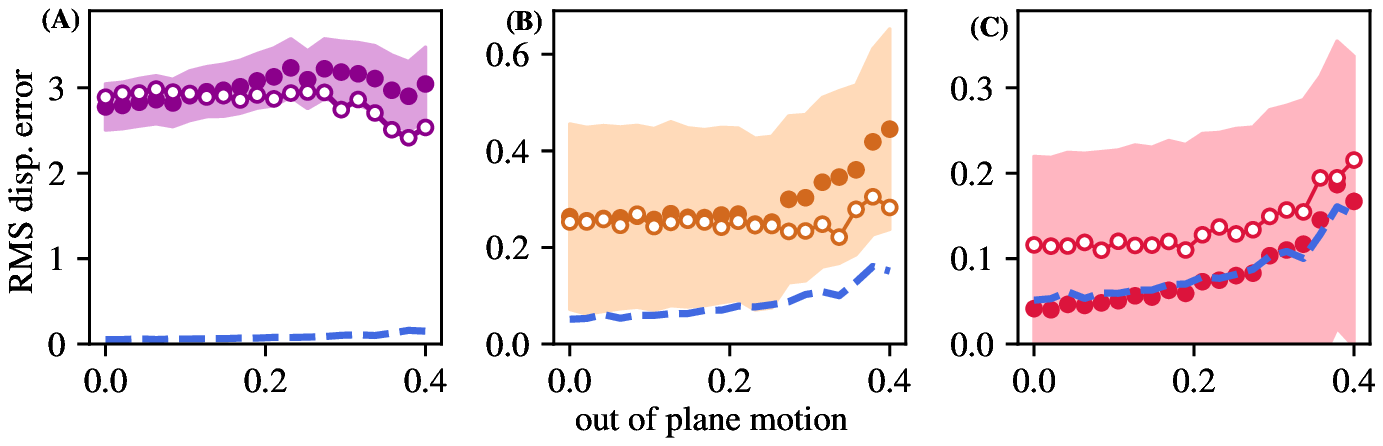}
    \caption{RMS error of particle displacements predicted by BCNNs and CNNs using interrogation windows only (IW, \textbf{A}), correlation maps and interrogation windows (CM+IW, \textbf{B}), and correlation maps only as input (CM, \textbf{C}) versus particle out of plane motion. Results from BCNNs are marked in solid circles with shaded regions representing standard deviations about the mean displacement prediction, while results from CNNs are marked in open circles. Each data point, solid or open, is an average of 100 samples. Shaded regions are standard deviations predicted by the BCNN, averaged over 100 samples. Blue dashed lines are the RMS error of predictions made by OpenPIV. We observe that as in Figure~\ref{fig:testset1_x}, the CM-BCNN and the CM-CNN (\textbf{C}) perform best out of the BCNNs and CNNs tested, respectively.}
    \label{fig:z}
\end{figure*}

\begin{figure*}
    \includegraphics{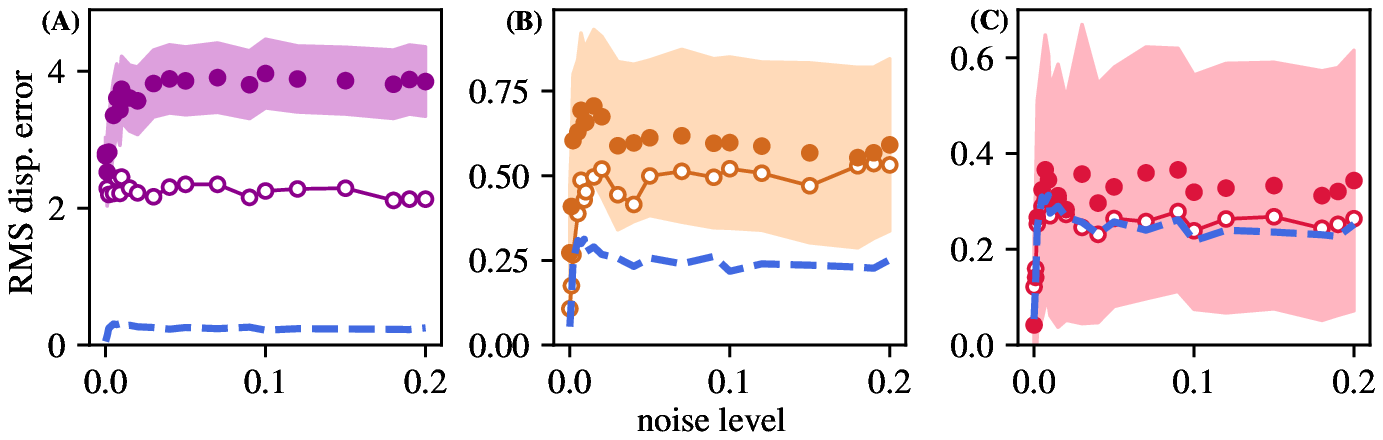}
    \caption{RMS error of particle displacements predicted by BCNNs and CNNs using interrogation windows only (IW, \textbf{A}), correlation maps and interrogation windows (CM+IW, \textbf{B}), and correlation maps only as input (CM, \textbf{C}) versus background noise level. Results from BCNNs are marked in solid circles with shaded regions representing standard deviations about the mean displacement prediction, while results from CNNs are marked in open circles. Each data point, solid or open, is an average of 100 samples. Shaded regions are standard deviations predicted by the BCNN, averaged over 100 samples. Blue dashed lines are the RMS error of predictions made by OpenPIV. We observe that as in Figure~\ref{fig:testset1_x}, the CM-BCNN and the CM-CNN (\textbf{C}) perform best out of the BCNNs and CNNs tested, respectively.}
    \label{fig:noise_level}
\end{figure*}

In Figure~\ref{fig:displacement}-\ref{fig:noise_level}, we show the root mean square (RMS) mean error of each BCNN's particle displacement prediction in solid circles and the RMS error of each CNN's particle displacement in hollow circles as a function of the error contributing parameters: particle displacement, particle density, particle diameter, particle shear, out of plane motion, and background noise level. Shaded regions indicate standard deviations from the RMS mean displacement error of each BCNN. Note that each standard deviation and each data point, solid or hollow, represents an average of 100 samples. Additionally, recall that the mean and standard deviation of each predicted particle displacement distribution generated by the BCNNs is estimated by drawing 2000 samples. For reference, similar analytic trends are demonstrated in Raffel \cite{Raffel1998}.

As observed previously on Test Set I, out of the 3 classes of BCNNs and CNNS tested, the BCNN and CNN using correlation maps only as inputs (CM-BCNN, CM-CNN) perform the best on Test Set II. The mean particle displacement predictions made by the CM-BCNN match or exceed the accuracy of those predicted by OpenPIV throughout ramps of particle shear (Figure~\ref{fig:shear}) and out of plane motion (Figure~\ref{fig:z}). 

During sweeps of particle displacement (Figure~\ref{fig:displacement}), particle density (Figure~\ref{fig:density}), and particle diameter (Figure~\ref{fig:diameter}), the mean displacement predictions made by the CM-BNN exceed or match those made by OpenPIV in accuracy for the majority of the error  contributing parameter range tested. As expected, the CM-BCNN performs poorly when presented with particle displacements outside of its training regime of $x, y \in [-4,4]$ pixels. Additionally, while ramping the error contributing parameter of background noise level (Figure~\ref{fig:noise_level}), the mean particle displacement predictions made by the CM-BCNN are less accurate but within $\sim 0.1$ pixels of those made by OpenPIV.

The standard deviations of predicted RMS particle displacements output by the BCNNs in Figure~\ref{fig:displacement}-\ref{fig:noise_level} range from from $0.23$ to $0.54$ for the IW-BCNN, from $0.17$ to $0.28$ for the CM+IW-BCNN, and from $0.17$ to $0.31$ for the CM-BCNN. Since the uncertainties output from the BCNN are computed by taking the standard deviation of $2000$ samples drawn from the BCNN, a smaller uncertainty indicates a narrower distribution of possible particle displacement predictions and a larger uncertainty indicates a wider distribution. Note that for all the BCNNs in Figure~\ref{fig:displacement}-\ref{fig:noise_level}, the minimum standard deviation of predicted RMS particle displacement seems to be around $0.2$ pixels. This uncertainty ``floor'' may be due to some limit to accuracy introduced by the neural networks themselves or the parameters used to synthesize Test Set II. This phenomenon merits investigation in future work.

It is important to again note the quantitative similarities between the results for each BCNN and its corresponding CNN (i.e. CM-BCNN and its deterministic version, CM-CNN) in Figure~\ref{fig:displacement}-\ref{fig:noise_level}. As discussed previously in {\em Results: Test Set I}, the reason for this similarity is that a BCNN acts as an ensemble of CNNs of identical architecture. This is particularly apparent in Figure~\ref{fig:displacement}-\ref{fig:z}, where all RMS displacement errors generated by CNN models fall within roughly 1 standard deviation of the RMS mean displacement error of their corresponding BCNN model. The only inconsistency in this trend can be found in Figure~\ref{fig:noise_level}, where the IW-CNN is not included within error bars of the IW-BCNN. This anomaly occurs due to the following: a BCNN represents an {\em ensemble} of CNNs, some of which may be more accurate or less accurate than $\pm 1$ standard deviation away from the mean output of the CNN ensemble. The IW-CNN in Figure~\ref{fig:noise_level} is one of these CNNs which lies outside of $\pm 1$ standard deviations from the mean BCNN output, as it lies at ~3 standard deviations from the mean curve.

\subsection{Performing full PIV using the CM-BCNN}
\label{sec:full}

\begin{table}
\centering
\caption{Parameters used to generate full image pair}
\label{tbl:full_piv}
  \begin{tabular}{ | p{6cm} | p{1.8cm} |}
    \hline
    Parameter & Value/range\\ \hline
    \hline
    Diameter of particles (pixels) & $3.0$ \\ \hline
    Particle density (particles/square pixel) & $0.098$ \\ \hline
    Strength of shear  (multiplier) & $0.0$  \\ \hline
    Angle of shear (degrees) & $0.0$ \\ \hline
    x-displacement (pixels) & $[-1.0,1.0]$ \\ \hline
    y-displacement (pixels) & $[-1.0,1.0]$ \\ \hline
    Out-of-plane displacement (pixels) & $0.0$ \\ \hline
    Noise level (fraction of standard deviation of image) & $0.0$  \\ \hline
  \end{tabular}
\end{table}

We have established that the CM-BCNN is the most accurate BCNN out of the three tested (IW, CM+IW, CM). We now demonstrate how the CM-BCNN can be used to generate accurate estimates of full flow fields from entire image frames. We simulate a simple flow field and generate two synthetic image frames using parameters shown in Table~\ref{tbl:full_piv}. 

\begin{figure*}
    \includegraphics{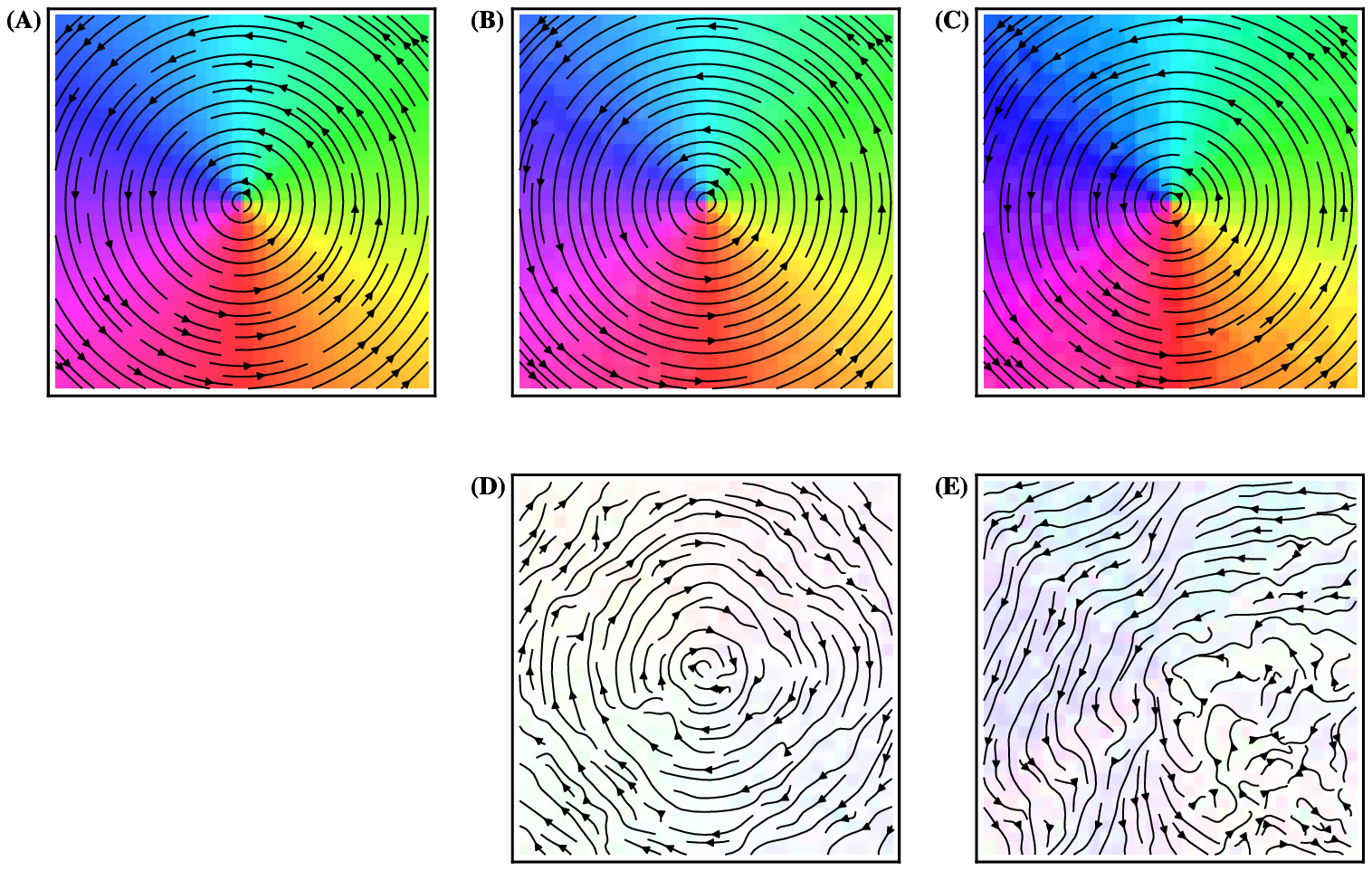}
    \caption{Performance of OpenPIV and CM-BCNN on a simulated test image pair with parameters listed in Table~. Color hue (i.e. blue, green, ...) indicates the direction of each interrogation window displacement vector, and color saturation indicates the magnitude of each displacement vector. Vectors with a smaller magnitude are less saturated, while vectors with a larger magnitude are more saturated. \textbf{(A)} shows the true flow field, \textbf{(B)} shows the flow field estimated by OpenPIV, and \textbf{(C)} shows the mean flow field estimated by the CM-BCNN. \textbf{(D)} and \textbf{(E)} show the errors of the OpenPIV generated flow field and the CM-BCNN generated flow field, respectively. Note that the errors generated by OpenPIV are roughly symmetric about both the x and y axes, while the errors generated by the CM-BCNN are not. Errors generated by OpenPIV (shown in \textbf{(D)}) fall in $[-0.20,0.13]$ in the x-direction and $[-0.16,0.11]$ in the y-direction. Errors generated by the CM-BCNN (shown in \textbf{(E)}) are bounded by $[-0.23, 0.13]$ in the x-direction and $[-0.24, 0.09]$ in the y-direction.}
    \label{fig:full_image}
\end{figure*}

In Figure~\ref{fig:full_image} we show the true flow field (Figure~\ref{fig:full_image}A), the flow field estimated by OpenPIV (Figure~\ref{fig:full_image}B), and the mean flow field predicted by the CM-BCNN (Figure~\ref{fig:full_image}C), from left to right. Different hues in Figure~\ref{fig:full_image} represent different flow vector direction, while color saturation represents flow vector magnitude. 

Slight variations from the true flow field in vector direction and magnitude can be observed in both the OpenPIV estimated flow field and the CM-BCNN estimated flow field (Figure~\ref{fig:full_image}D, Figure~\ref{fig:full_image}E, respectively). Note that the errors generated by OpenPIV are roughly symmetric and structured about both the x and y axes, while the errors generated by the CM-BCNN are not. This may be due to a systematic bias in the synthetic training data which might be rectified by additional training data. However, the error magnitude is comparable between OpenPIV and the CM-BCNN.

\begin{figure}
    \includegraphics{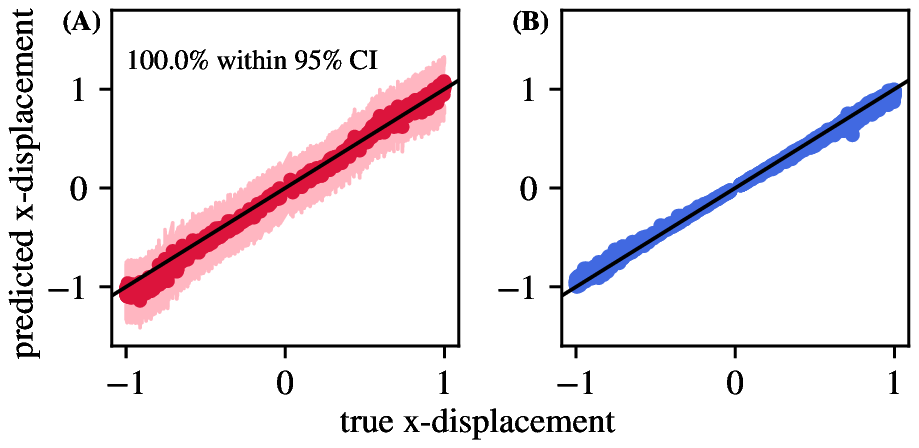}
    \caption{\textbf{(A)} Particle displacements within the flow field shown in Figure~\ref{fig:full_image} predicted by the CM-BCNN versus true particle displacements within the flow field shown in Figure~\ref{fig:full_image} in the $x$-direction. Note that $100 \%$ of the true particle displacements are contained within the $95 \%$ confidence interval predicted by the CM-BCNN. Shaded regions are standard deviations. \textbf{(B)} Particle displacements within the flow field shown in Figure~\ref{fig:full_image} predicted by OpenPIV vs true particle displacements in the x-direction. Results for the equivalent of \textbf{(A)} and \textbf{(B)} in the y-direction are similar and are shown in the Appendix~\ref{app:appendix} (Figure~\ref{fig:full_y}).}
    \label{fig:full_x}
\end{figure}

The CM-BCNN performs well in capturing the true flow upon visual inspection, with performance comparable to that of OpenPIV. In Figure~\ref{fig:full_x} and Figure~\ref{fig:full_y} we show the particle displacements predicted by OpenPIV and the CM-BCNN versus the true particle displacements between the image pair whose true flow field is shown in Figure~\ref{fig:full_image}A. The CM-BCNN is able to capture $100 \%$ of the true particle displacements within its $95\%$ confidence interval (Figure~\ref{fig:full_x}A, Figure~\ref{fig:full_y}A). The accuracy of the CM-BCNN is similar to that of OpenPIV, whose particle displacement predictions are shown in Figure~\ref{fig:full_x}B. 
\subsection{CM-BCNN generalizability}
\label{sec:2pass}

In the previous sections, we have demonstrated that the CM-BCNN is an effective tool for simultaneously producing accurate particle displacement predictions and quantifying their uncertainty. A tempting use for the CM-BCNN is to apply it to correlation maps generated by existing, perhaps state-of-the-art PIV codes. It is therefore important to determine  whether 1) the CM-BCNN is sufficiently general to estimate particle displacements and uncertainties from correlation maps originating from existing PIV algorithms and 2) if the CM-BCNN is sufficiently general, whether it is practical to do so. In this section, we test CM-BCNN's ability to generalize to existing PIV algorithms by evaluating its performance on correlation maps generated by a simple multi-pass algorithm. The simplest case of a multi-pass PIV algorithm is a double pass algorithm. This algorithm follows the steps enumerated in Table~\ref{tbl:2pass}.

\begin{table}
\centering
\caption{Double pass PIV algorithm}
\label{tbl:2pass}
  \begin{tabular}{ | p{0.7cm} | p{7.5cm} | }
    \hline
    Step &  Operation \\ \hline
    \hline
    1 & Select \emph{Interrogation Window A} and \emph{Interrogation Window B} on the sequential image pair \emph{Image Frame A} and \emph{Image Frame B}. During the first pass, \emph{Interrogation Window A} and \emph{Interrogation Window B} are centered about the same coordinates on both \emph{Image Frame A} and \emph{Image Frame B}. \\ \hline
    2 & Compute the cross-correlation map of \emph{Interrogation Window A} and \emph{Interrogation Window B} and extract an integer estimate of the particle displacement between them via OpenPIV.\\ \hline
    3 & Shift \emph{Interrogation Window B} by the integer estimate of particle displacement calculated in 2). Compute the cross-correlation map of \emph{Interrogation Window A} and \emph{Interrogation Window B} and again extract the particle displacement between them via OpenPIV. \\ \hline
    4 & Add the particle displacements found in 2) and 3) together, yielding the final estimate of particle displacement. \\ \hline
  \end{tabular}
\end{table}

\begin{table}
\centering
\caption{Double pass CM-BCNN algorithm}
\label{tbl:2pass_bcnn}
  \begin{tabular}{ | p{0.7cm} | p{7.5cm} | }
    \hline
    Step &  Operation \\ \hline
    \hline
    1 & Select \emph{Interrogation Window A} and \emph{Interrogation Window B} on the sequential image pair \emph{Image Frame A} and \emph{Image Frame B}. During the first pass, \emph{Interrogation Window A} and \emph{Interrogation Window B} are centered about the same coordinates on both \emph{Image Frame A} and \emph{Image Frame B}. \\ \hline
    2 & Compute the cross-correlation map of \emph{Interrogation Window A} and \emph{Interrogation Window B} and extract an integer estimate of the particle displacement between them via OpenPIV. \\ \hline
    3 & Shift \emph{Interrogation Window B} by the integer estimate of particle displacement calculated in 2). Compute the cross-correlation map of \emph{Interrogation Window A} and \emph{Interrogation Window B} and extract the particle displacement between them via the CM-BCNN, yielding a mean and standard deviation prediction. \\ \hline
    4 & Use the CM-BCNN to extract a mean and standard deviation particle displacement from the correlation map calculated in 2). Discard the non-integer part of the mean particle displacement prediction. \\ \hline
    5 & Add the particle displacements found in 4) and 3) together, yielding the final estimate of particle displacement. Propagate the uncertainties found in 4) and 3) to obtain the uncertainty of the final particle displacement prediction.  \\ \hline
  \end{tabular}
\end{table}

In a typical multi-pass PIV algorithm, steps 2) and 3) in Table~\ref{tbl:2pass} are iterated many times such that the final estimate of particle displacement converges to a stable value. Additionally, a typical multi-pass PIV algorithm involves removal of spurious displacements, or large errors in predicted particle displacement, in step 2). We accomplish this by defaulting to a single pass algorithm for particle displacements estimated in step 2) greater than $10$ pixels. 

Many widely used PIV codes \cite{Scarano2000, Stanislas2008} utilize either multi-pass or multi-grid methods, both which involve iterative cross-correlation of interrogation window pairs, shifting an interrogation window by the estimated particle displacement found in the last algorithm iteration. This is because multi-pass PIV algorithms have been shown to reduce errors in estimated particle displacement by optimizing the number of particles contained within each interrogation window \cite{Adrian2011, Raffel1998}. 

Additionally, we note that window deformation is another commonly used method in popular PIV codes: we did not implement it in this work, as window deformation also requires the iterative use of interrogation windows offset by estimated particle displacements \cite{Huang1993}. As a baseline test of whether correlation maps generated by PIV algorithms involving interrogation window offsets can be analyzed by the CM-BCNN, we implement the double pass algorithm enumerated in  Table~\ref{tbl:2pass} using both the CM-BCNN and OpenPIV for comparison.

\begin{figure}
    \includegraphics{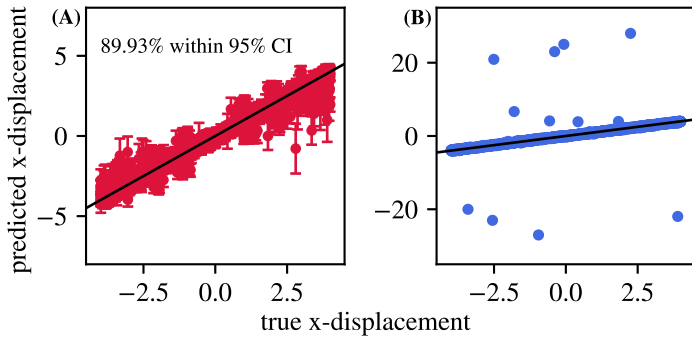}
    \caption{\textbf{(A)} Particle displacements predicted by the CM-BCNN using a double pass algorithm versus true particle displacements in the $x$-direction. The \emph{double pass} algorithm used with the CM-BCNN is detailed in Table~\ref{tbl:2pass_bcnn}. Although not as accurate as the single pass CM-BCNN shown in Figure~\ref{fig:testset1_x}, the double pass CM-BCNN implementation shows good accuracy on Test Set I, suggesting that the CM-BCNN is robust enough to generalize to correlation maps generated via multiple algorithms. \textbf{(B)} Particle displacements predicted by OpenPIV using a double pass algorithm vs true particle displacements in the $x$-direction. Results for the equivalent of \textbf{(A)} and \textbf{(B)} in the $y$-direction are similar and are shown in Appendix~\ref{app:appendix} (Figure~\ref{fig:2pass_y}).}
    \label{fig:2pass_x}
\end{figure}

First, we implement the double pass algorithm described in Table~\ref{tbl:2pass} using OpenPIV and run this algorithm on Test Set I. In Figure~\ref{fig:2pass_x}B, we show the results of this calculation for displacements in the x-direction. The double pass PIV algorithm performs slightly better than the single pass algorithm on Test Set I, with fewer spurious displacements, and its predicted vs true particle displacement has an $R^2$ value of $0.723$ as opposed to the $R^2$ of $0.722$ found for the single pass equivalent (Figure~\ref{fig:testset1_x}G). Results for displacements in the y-direction are similar and are shown in Appendix~\ref{app:appendix} (Figure~\ref{fig:2pass_y}B): the predicted vs true particle displacement curve for the double pass algorithm has an $R^2$ value of $0.8599$ while the single pass equivalent has an $R^2$ value of $0.8596$.

Next, we use the CM-BCNN to perform the double-pass algorithm. Note that we still use OpenPIV to calculate the interrogation window offsets. We choose to do this because when implementing the CM-BCNN to simultaneously provide particle displacement estimates and uncertainties from correlation maps calculated via an existing algorithm, it is more practical to not make any modifications to the existing algorithm and simply introduce the CM-BCNN afterwards. 

In Table~\ref{tbl:2pass_bcnn}, we outline the double pass algorithm used with the CM-BCNN, step by step. We run this algorithm on Test Set I and show the results of this calculation for displacements in the x-direction in Figure~\ref{fig:2pass_x}A. The double pass CM-BCNN captures $89.93 \%$ of true particle displacements in its $95\%$ confidence interval, a result not as accurate as that obtained by the single pass CM-BCNN, but still sufficient to suggest that the CM-BCNN can indeed be generalized to correlation maps generated through multi-pass algorithms, with some sacrifice of accuracy. Results for displacements in the y-direction are similar and are shown in Appendix~\ref{app:appendix} (Figure~\ref{fig:2pass_y}A). 

\begin{figure}
    \includegraphics{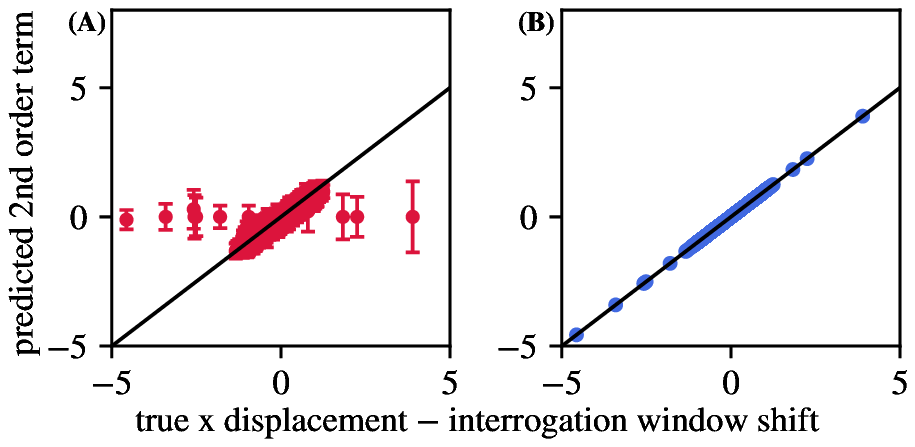}
    \caption{\textbf{(A)} The 2nd order term of the predicted particle displacement on Test Set I generated by step 4) of the double pass CM-BCNN algorithm detailed in Table~\ref{tbl:2pass_bcnn} versus the true 2nd order particle displacement term: the difference between the true x-displacement and the interrogation window shift generated by step 2) of the algorithm described in Table~\ref{tbl:2pass_bcnn}. Error bars are standard deviations. \textbf{(B)} The OpenPIV equivalent of \textbf{(A)}: the 2nd order term of the predicted particle displacement on Test Set I generated by step 3) of the double pass OpenPIV algorithm detailed in Table~\ref{tbl:2pass} versus the true 2nd order particle displacement term: the difference between the true x-displacement and the interrogation window shift generated by step 2) of the algorithm described in Table~\ref{tbl:2pass}. Note that the CM-BCNN struggles to match the accuracy of OpenPIV on correlation maps generated from multi-pass algorithms, such as those used to generate this figure. Results for the y-direction are similar and are shown in Appendix~\ref{app:appendix} (Figure~\ref{fig:2pass_error_y}).}
    \label{fig:2pass_error_x}
\end{figure}

This lack of accuracy is surprising, as one would expect that the double pass CM-BCNN would be more accurate than the single pass CM-BCNN, as was the case for the OpenPIV single and double pass algorithms. The reason for double pass CM-BCNN's inaccuracy compared to its single pass equivalent is found in Figure~\ref{fig:2pass_error_x}. In Figure~\ref{fig:2pass_error_x}A, we plot the 2nd order term of the predicted particle displacement on Test Set I generated by step 4) of the double pass CM-BCNN algorithm detailed in Table~\ref{tbl:2pass_bcnn} versus the true 2nd order particle displacement term: the difference between the true x-displacement and the interrogation window shift generated by step 2) of the algorithm described in Table~\ref{tbl:2pass_bcnn}. In Figure~\ref{fig:2pass_error_x}B, we plot the OpenPIV equivalent of Figure~\ref{fig:2pass_error_x}A. Note that the double pass CM-BCNN struggles to make accurate second order displacement predictions on Test Set I compared to the single pass CM-BCNN's performance on Test Set I \emph{and} compared to the double pass OpenPIV algorithm's performance shown in Figure~\ref{fig:2pass_error_x}A. This is because the CM-BCNN is not trained on correlation maps generated by double pass algorithms. Results for particle displacements in the y direction are similar and can be found in Appendix~\ref{app:appendix} (Figure~\ref{fig:2pass_error_y}).

\begin{figure*}
    \includegraphics{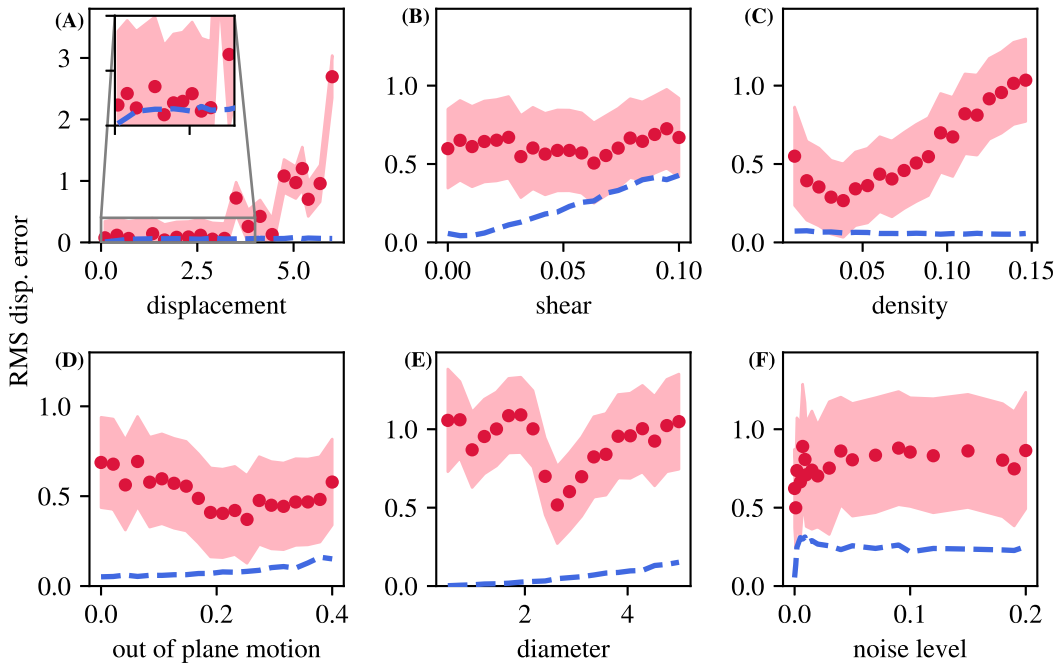}
    \caption{RMS displacement error for the CM-BCNN using a double pass algorithm versus error contributing parameters: RMS particle displacement \textbf{(A)}, particle shear \textbf{(B)}, particle density \textbf{(C)}, out of plane motion \textbf{(D)}, particle diameter \textbf{(E)}, and background noise level \textbf{(F)}. Comparing these results to their single-pass CM-BCNN equivalents (rightmost column, Figures~\ref{fig:displacement}-\ref{fig:noise_level}), the double pass CM-BCNN is far less accurate, especially in response to ramping the error contributing parameters of shear, density, out of plane motion, and diameter. The double pass CM-BCNN is less accurate than its single pass equivalent because the CM-BCNN was trained on single pass correlation maps, not double pass correlation maps. Justification for this claim can be found in Figure~\ref{fig:2pass_error_x}, where we show that the 2nd order displacement term estimated by the CM-BCNN is far less accurate than its 1st order estimates.}
    \label{fig:2pass_sweeps} 
\end{figure*}

The double pass CM-BCNN's performance on Test Set II is also subpar compared to that of the single pass BCNN, as shown in Figure~\ref{fig:2pass_sweeps}. The double pass CM-BCNN's lack of accuracy on Test Set II compared to the single pass CM-BCNN also emerges because the CM-BCNN is not introduced to correlation maps generated by double pass algorithms during training. However, the CM-BCNN still is able to consistently make particle displacement predictions with errors in RMS mean displacement below 1 pixel for most ranges of error contributing parameters tested, implying that although some accuracy was lost, the CM-BCNN is still able to recognize features of correlation maps generated by double pass algorithms.

We conclude that with a moderate loss in accuracy, the CM-BCNN generalizes successfully to correlation maps generated by multi pass algorithms. In future work, techniques such as augmenting the training set (i.e. rotating and adding background noise to training data), fine-tuning the CM-BCNN with correlation maps produced by multi-pass algorithms, and/or including multi-pass correlation maps in the training data could be implemented to decrease the loss in accuracy observed upon applying the CM-BCNN to multi-pass correlation maps.

We end our discussion of the application of the CM-BCNN to multi-pass algorithms by noting a mathematical consideration. Since the uncertainty of each particle displacement prediction made on each correlation map must be propagated to yield the final particle displacement prediction uncertainty as described in step 5) of the double pass CM-BCNN algorithm detailed in Table~\ref{tbl:2pass_bcnn}, with every additional pass the final particle displacement uncertainty grows monotonically. Thus, if the spread is large on particle displacement distributions predicted by the CM-BCNN for a given correlation map, it may be impractical to implement a multi-pass algorithm.

\section{Discussion}
In this work, for the first time, we have shown that Bayesian convolutional neural networks are effective for simultaneous PIV calculation and uncertainty quantification. First, we found that correlation maps (CM) are the best inputs for both our BCNN and CNN models out of the inputs tested (IW, CM+IW, and CM). Furthermore, by testing each of our neural networks on Test Set I and Test Set II, we demonstrated that the CM-BCNN is able to both output accurate predictions rivaling OpenPIV and provide simultaneous uncertainty quantification. Upon application to Test Set I, it was seen that CM-BCNN predicted uncertainties (95\% confidence intervals) capture $\approx99\%$ of true displacements. Additionally, via testing on Test Set II, we observed that the CM-BCNN is fairly robust to ramping error contributing parameters. Finally, the quantitative similarity between BCNN and corresponding CNN predictions across tests on Test Sets I and II are consistent with the theoretical expectation that predictions from a given CNN should fall within the distribution of predictions generated by a BCNN of identical architecture. 

Once we established that the CM-BCNN was the best performing neural network out of those tested, we applied it to the test problem of a full synthetic image pair with a simple flow field. The CM-BCNN successfully captured $100\%$ of true particle displacements in its $95\%$ confidence interval, with accuracy comparable to that of OpenPIV. As a final test, we evaluated the generalizability of the CM-BCNN to correlation maps generated by multi-pass algorithms. With a moderate loss of accuracy, we found that the CM-BCNN was able to generalize to the multi-pass correlation maps. By adjusting training datasets and finetuning the CM-BCNN, future work may yield a CM-BCNN which can be applied to both single and multi pass algorithms with no loss of accuracy.

The network architectures used in this work are very simple, lending themselves to the application of interpretability studies to understand the features in correlation maps being used for inference of particle displacement. However, it is always possible in deep ML research that modifications in architecture or variations in the training setup could yield more accurate results. Further work on exhaustive studies of the BCNN approach to PIV analysis can and should be carried out. One indication that there may be future improvements in this application is the universally poor performance of BCNNs and CNNs trained on just the flow field interrogation windows. The information to compute correlation maps is certainly contained in the interrogation windows and therefore should be able to be utilized by the CNNs. However, the transformation from an interrogation window to a correlation map is a very specific non-linear transformation and therefore the simple architectures explored here may be insufficient. A future direction of research may be to include more complex structures in the BCNN architecture, i.e. the pyrimidal feature extraction used in state-of-the-art CNNs such as LiteFlowNet \cite{LiteFlowNet2018}. 

Besides improving the network architecture, application of deep ML to PIV analysis opens many avenues for inferring fluid flow information. In many applications there are multiple flow diagnostics. It is a simple task to construct networks that can take in highly varied sets of simultaneous diagnostic input. Understanding how to construct effective, stable, network architectures which couple multiple diagnostics is an exciting area of future research.

Finally, we stress the novelty and utility of this work. By using BCNNs to perform PIV tasks, we have demonstrated that predicted particle displacements and their uncertainties can be generated simultaneously from the same algorithm. We emphasize that the uncertainties produced from the BCNNs are \emph{directly} produced from the algorithm that generates particle displacement predictions, which is not the case for many existing PIV uncertainty quantification methods. We suggest that using BCNNs to perform PIV is a solution to the absence of a universally accepted uncertainty quantification method for classical and CNN-based PIV algorithms, and anticipate that understanding how to apply BCNNs to PIV will become a compelling field of study.

\section*{Acknowledgements}{\par\addvspace{12pt}}
    This work was supported by the U.S. Department of Energy through the Los Alamos National Laboratory. Los Alamos National Laboratory is operated by Triad National Security, LLC, for the National Nuclear Security Administration of U.S. Department of Energy (Contract No. 89233218CNA000001). This work has been approved for release under LA-UR-20-29836.
%\begin{acknowledgements}
%Thanks to people and grants
%\end{acknowledgements}
\bibliographystyle{unsrt}
\bibliography{bib}
\renewcommand{\thefigure}{A\arabic{figure}}
\setcounter{figure}{0}

\appendix 
\section{Additional figures}
\label{app:appendix}

\begin{figure}
    \includegraphics{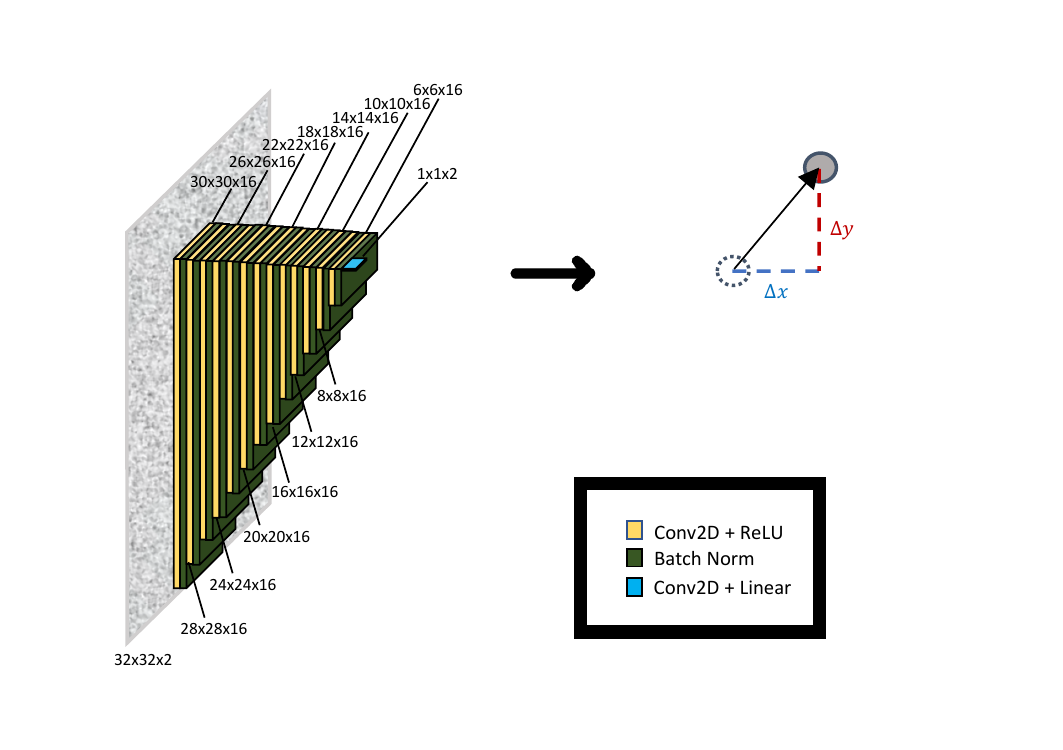}
    \caption{Structure of the neural networks with interrogation windows only used as inputs (IW). All neural networks used in this paper follow the same basic structure: repeated application of convolutional layers followed by batch normalization layers until the input is decreased in dimension to that of a 2D vector.}
    \label{fig:iw-bcnn}
\end{figure}

\begin{figure}
    \includegraphics{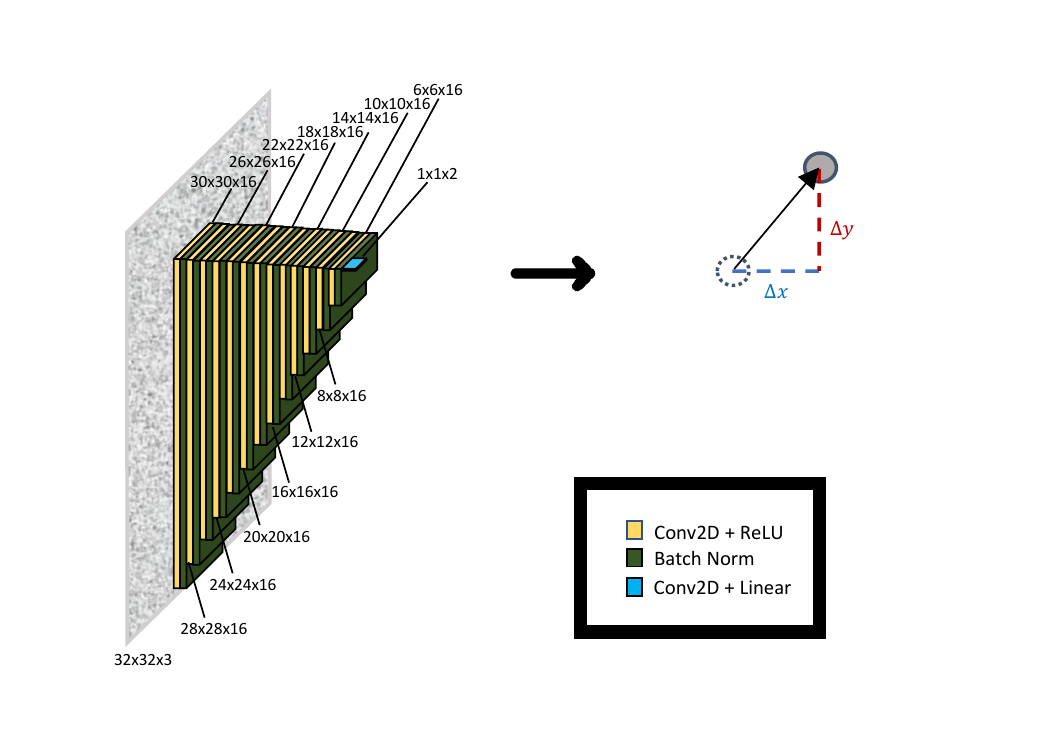}
    \caption{Structure of the neural networks with correlation maps and interrogation windows used as inputs (CM+IW). All neural networks used in this paper follow the same basic structure: repeated application of convolutional layers followed by batch normalization layers until the input is decreased in dimension to that of a 2D vector.}
    \label{fig:cm+iw-bcnn}
\end{figure}

In this section, we provide figures relevant to the main text but not included. First, we show the equivalents of Figure~\ref{fig:cm-bcnn} for the IW and CM+IW neural networks, found in Figure~\ref{fig:iw-bcnn} and Figure~\ref{fig:cm+iw-bcnn}, respectively. 

\begin{figure*}
    \includegraphics{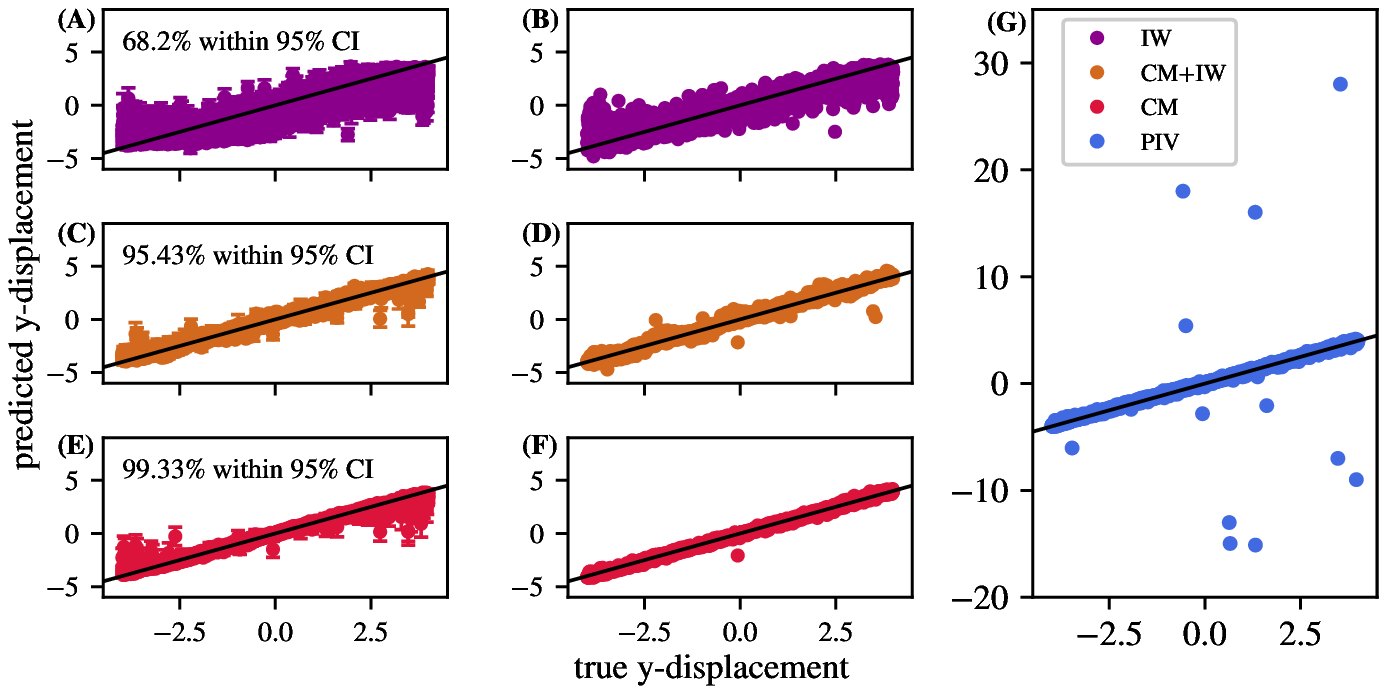}
    \caption{ Predicted particle displacements vs true displacements from Test Set I in the y direction. The equivalent plot in the x-direction, Figure~\ref{fig:testset1_x} is similar and is found in the main text. \textbf{(A, C, E)} show particles displacements predicted by Bayesian neural networks with interrogation windows only as inputs (IW-BCNN, purple), correlation maps and interrogation windows as inputs (CM+IW-BCNN, brown), and correlation maps only as inputs (CM-BCNN, red), respectively vs true particle displacements. \textbf{(B, D, F)} show particle displacements predicted by the deterministic versions of \textbf{(A, C, E)}, respectively vs true particle displacements (IW-CNN, CM+IW-CNN, and CM-CNN). \textbf{(F)} shows particle displacements predicted by OpenPIV (PIV, blue) vs true particle displacements. As in Figure~\ref{fig:testset1_x}, we observe that the CM-BCNN and the CM-CNN perform the best out of the BCNNs and CNNs tested, respectively. All of the tested BCNNs and CNNs decrease the number of large errors in displacement, or spurious displacements, compared to the OpenPIV algorithm \textbf{(F)}.}
    \label{fig:testset1_y}
\end{figure*}

\begin{figure}
    \includegraphics{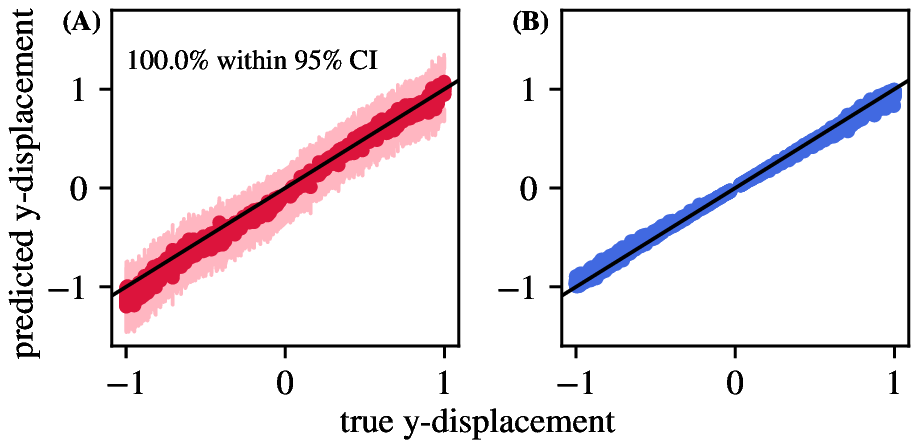}
    \caption{\textbf{(A)} Particle displacements within the flow field shown in Figure~\ref{fig:full_image} predicted by the CM-BCNN vs true particle displacements within the flow field shown in Figure~\ref{fig:full_image} in the y-direction. Shaded regions are standard deviations. Note that $100 \%$ of the true particle displacements are contained within the $95 \%$ confidence interval predicted by the CM-BCNN. \textbf{(B)} Particle displacements within the flow field shown in Figure~\ref{fig:full_image} predicted by OpenPIV vs true particle displacements in the y-direction. Results for the equivalent of \textbf{(A)} and \textbf{(B)} in the x-direction are similar and are shown in the main text (Figure~\ref{fig:full_x}).}
    \label{fig:full_y}
\end{figure}

\begin{figure}
    \includegraphics{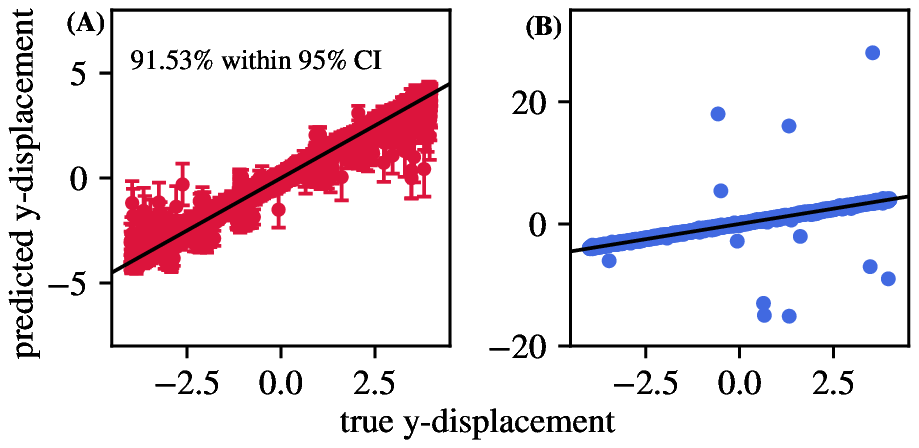}
    \caption{\textbf{(A)} Particle displacements predicted by the CM-BCNN (the BCNN using correlation maps only as inputs) using a double pass algorithm vs true particle displacements in the y-direction. The ``double pass'' algorithm used with the CM-BCNN involves 1) estimating particle displacement via OpenPIV, 2) shifting the latter image frame's interrogation window by the particle displacement estimate, and 3) calculating the correlation map of the new interrogation window pair. This yields 2 correlation maps: one from the first pass and another from the second pass. Both of these correlation maps are fed into the BCNN to predict particle displacement in the ``double pass'' algorithm. Although not as accurate as the ``single pass'' CM-BCNN shown in Figure~\ref{fig:testset1_x}, the double pass CM-BCNN implementation shows good accuracy on Test Set 1, suggesting that the CM-BCNN is robust enough to generalize to correlation maps generated via multiple algorithms. \textbf{(B)} Particle displacements predicted by the OpenPIV using a double pass algorithm vs true particle displacements in the x-direction. Results for the equivalent of \textbf{(A)} and \textbf{(B)} in the x-direction are similar and are shown in the main text (Figure~\ref{fig:2pass_x}).}
    \label{fig:2pass_y}
\end{figure}

\begin{figure}
    \includegraphics{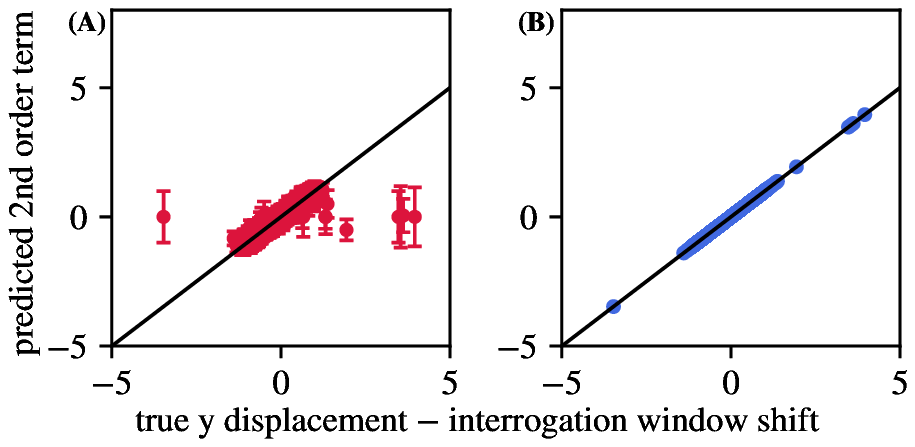}
    \caption{\textbf{(A)} The 2nd order term of the predicted particle displacement on Test Set I generated by step 4) of the double pass CM-BCNN algorithm detailed in Table~\ref{tbl:2pass_bcnn} versus the true 2nd order particle displacement term: the difference between the true y-displacement and the interrogation window shift generated by step 2) of the algorithm described in Table~\ref{tbl:2pass_bcnn}. Error bars are standard deviations. \textbf{(B)} The OpenPIV equivalent of \textbf{(A)}: the 2nd order term of the predicted particle displacement on Test Set I generated by step 3) of the double pass OpenPIV algorithm detailed in Table~\ref{tbl:2pass} versus the true 2nd order particle displacement term: the difference between the true y-displacement and the interrogation window shift generated by step 2) of the algorithm described in Table~\ref{tbl:2pass}. Note that the CM-BCNN struggles to match the accuracy of OpenPIV on correlation maps generated from multi-pass algorithms, such as those used to generate this figure. Results for the y-direction are similar and are shown in the main text  (Figure~\ref{fig:2pass_error_x}).}
    \label{fig:2pass_error_y}
\end{figure}

Next, we show results in the y-direction for the plots shown exclusively in the x-direction in the main text. This includes the y-direction equivalents to Figure~\ref{fig:testset1_x}, Figure~\ref{fig:full_x}, Figure~\ref{fig:2pass_x}, and Figure~\ref{fig:2pass_error_x}, which are shown in Figure~\ref{fig:testset1_y}, Figure~\ref{fig:full_y}, Figure~\ref{fig:2pass_y}, and Figure~\ref{fig:2pass_error_y}, respectively.

\section{Synthetic image generation}
\label{app:imagegen}
Synthetic particle image pairs are generated by uniformly seeding an initial image with particle locations and then updating these locations to create the second image in the pair. Particle locations $(x[t], y[t], z[t])$ are updated using the following equation of motion: 
\begin{equation}
    x[t]=x[t-1]+\Delta x+ \big(-X[t-1]\sin\alpha_g+Y[t-1]\cos\alpha_g\big)g\cos\alpha_g
\end{equation}
\begin{equation}
    y[t]=y[t-1]+\Delta y+ \big(-X[t-1]\sin\alpha_g+Y[t-1]\cos\alpha_g\big)g\sin\alpha_g
\end{equation}
\begin{equation}
    z[t]=z[t-1] + \Delta z
\end{equation}
where $(x[t], y[t], z[t])$ is the particle's updated position at time $t$, $(\Delta x, \Delta y, \Delta z)$ is the particle displacement vector, and $g$ is a multiplier which tunes the magnitude of shear displacement with direction $\alpha_g$. The coordinates $(X[t-1],Y[t-1])$ are a transformation of the image coordinates $(x[t-1],y[t-1])$ and represent the particle's distance from the center of the image minus the overlap term $lw$, where $l=0.75$ is a multiplier of interrogation window size $w$. For each image frame in the image pair, each synthetic particle is given a Gaussian intensity profile. 

\section{Negative log-likelihood is equivalent to MSE for Gaussian distributions with fixed standard deviation}
\label{app:math}

Here we show that for Gaussian $ p(y_i|\mathbf{x}_i, \mathbf{w}_i)$ with fixed standard deviation $s_i = s$, the negative log likelihood is interchangeable with the mean squared error of data $\mathcal{D}$.

Suppose $ p(y_i|\mathbf{x}_i, \mathbf{w}_i)$ is a Gaussian
\begin{equation}
    p(y_i|\mathbf{x}_i, \mathbf{w}_i) \propto e^{-\frac{(y_i-m_i)^2}{2s_i^2}}.
\end{equation}
Then the log likelihood function is
\begin{equation}
    \log p(\mathcal{D}|\mathbf{w}) = \log \prod_{i} p(y_i|\mathbf{x}_i, \mathbf{w}_i) = C - \sum_i \frac{(y_i-m_i)^2}{2s_i^2},
\end{equation}
where $C$ is a constant.
Assuming that the standard deviation of each $ p(y_i|\mathbf{x}_i, \mathbf{w}_i)$ is equal, 
\begin{equation}{\label{eq:msell}}
    \log p(\mathcal{D}|\mathbf{w}) = C -\frac{1}{2 s^2}\sum_i (y_i-m_i)^2.
\end{equation}
Plugging Eq.~\ref{eq:msell} into Eq.~\ref{eq:costapprox}, we can drop the constant $C$ since Eq.~\ref{eq:costapprox} will be minimized:
\begin{equation}{\label{eq:costplug}}
\mathcal{F}(\mathcal{D}, \boldsymbol{\theta})  \approx \frac{1}{N}\sum_{j=1}^{N}[ \log q(\mathbf{w}^{(j)}|\boldsymbol{\theta})-\log p(\mathbf{w}^{(i)})] + \frac{1}{2 s^2}\sum_i (y_i-m_i)^2.
\end{equation}
Note that the second term of Eq.\ref{eq:costplug} is proportional to the mean squared error between data $\{y_i\}$ and mean $\{m_i\}$:
\begin{equation}
    \text{MSE} = \frac{1}{n}\sum_{i=1}^{n} (y_i-m_i)^2.
\end{equation}
Thus, we can replace the second term in Eq.~\ref{eq:costplug} with the following:
\begin{equation}
    \mathcal{F}(\mathcal{D}, \boldsymbol{\theta})  \approx \frac{1}{N}\sum_{j=1}^{N}[ \log q(\mathbf{w}^{(j)}|\boldsymbol{\theta})-\log p(\mathbf{w}^{(i)})] + \text{MSE}.
\end{equation}
\end{document}